\documentclass[aps,prx,twocolumn,superscriptaddress,showpacks]{revtex4-1}
\usepackage{amsfonts}
\usepackage{amsmath}
\usepackage{amssymb}
\usepackage{graphicx}
\usepackage{epstopdf}
\usepackage{color}
\usepackage{bbold}

\renewcommand{\vec}[1]{\boldsymbol{#1}}

\begin{document}

\title{Control of Spin Diffusion and Suppression of the Hanle Effect \\  by the Coexistence of Spin and Valley Hall Effects}
\author{Xian-Peng Zhang}
\affiliation{Donostia International Physics Center (DIPC), Manuel de
Lardizabal, 4. 20018, San Sebastian, Spain}
\affiliation{Centro de Fisica de Materiales (CFM-MPC), Centro Mixto CSIC-UPV/EHU,
20018 Donostia-San Sebastian, Basque Country, Spain}

\author{Chunli Huang}
\affiliation{Department of Physics, The University of Texas at Austin, Austin, Texas 78712, USA}

\author{Miguel A. Cazalilla}
\affiliation{Donostia International Physics Center (DIPC), Manuel de
Lardizabal, 4. 20018, San Sebastian, Spain}
\affiliation{Department of Physics, National Tsing Hua University, Hsinchu 30013,
Taiwan}
\affiliation{National Center for
Theoretical Sciences (NCTS), Hsinchu 30013, Taiwan}

\begin{abstract}
In addition to spin, electrons in many materials possess an additional pseudo-spin degree of freedom known as `valley'. In materials where the spin and valley degrees of freedom are weakly coupled, they can be both excited and controlled independently. In this work, we study a model describing the interplay of the spin and valley Hall effects in such two-dimensional materials. We demonstrate the emergence of an additional longitudinal neutral current that is both spin and valley polarized. The additional neutral current allows to control the spin density by tuning the magnitude of the valley Hall effect.  In addition, the interplay of the two effects can suppress the Hanle effect, that is, the oscillation of the nonlocal resistance of a Hall bar device with in-plane magnetic field. The latter observation  provides a possible explanation for the absence of the Hanle effect in a number of recent experiments. Our work also opens the possibility to engineer the conversion between the valley and spin degrees of freedom in two-dimensional materials.
\end{abstract}

\maketitle

\textit{Introduction:} 
Spin-orbitronics ~\cite {xiao2010berry,sinova2015spin,Nagaosa_RevModPhys.82.1539,sinova2004universal,huertas2006spin}  and valleytronics ~\cite {xu2014spin,cao2012valley,rycerz2007valley,sie2015valley} aim at manipulating internal degrees of freedom of Bloch electrons, which can have applications in low-energy consumption electronics and quantum computation. Some two-dimensional (2D) materials  such as transition metal dichalcogenides (TMD) ~\cite {zhu2011giant,zhang2014direct} are known to exhibit large spin-orbit coupling (SOC), whilst for others like graphene, it has been predicted that SOC can be enhanced by means of decoration with various types of absorbates  \cite{neto2009impurity,gmitra2013spin,irmer2015spin,ding2011engineering}  or by proximity to a substrate such as a TMD material \cite{gmitra2015graphene,wang2015strong,cummings2017giant,garcia2017spin}. Both intrinsic and extrinsic SOC can lead to the spin Hall effect (SHE), i.e. the generation of a spin current perpendicular to the applied electric  field.

 In many 2D materials Bloch electrons are endowed with an additional pseudo-spin degree of freedom known as `valley'. The latter is related to the existence of independent high symmetry points in the Brillouin zone where the band structure exhibits degenerate Dirac points or extrema~\cite
{katsnelson2012graphene,neto2009electronic,nebel2013valleytronics,schaibley2016valleytronics}. Analogous to the SHE, these systems are capable of exhibiting the so-called valley Hall effect (VHE) \cite{xiao2007valley,gorbachev2014detecting,beconcini2016nonlocal,zhang2016valley}, i.e. the appearance of a  transverse valley-polarized  bulk current in response to the application of an external electric field. Indeed, symmetry considerations imply that spin and valley are coupled in  materials with broken spin-rotation and/or inversion symmetry. As such, 2D materials and van der Walls heterostructures  have emerged as some of the most promising platforms to investigate this interesting interplay of spintronics and valleytronics. While spin and valley currents are electrically neutral, both currents carry angular momentum. In pristine graphene where SOC is negligible,  valley current carries  orbital angular momentum while spin current carries spin angular momentum. In the opposite limit,  in TMDs, for which the spin-momentum locking SOC is strong, there is often no distinction between the two.

\begin{figure}[t]
\begin{center}
\includegraphics[width=0.42\textwidth]{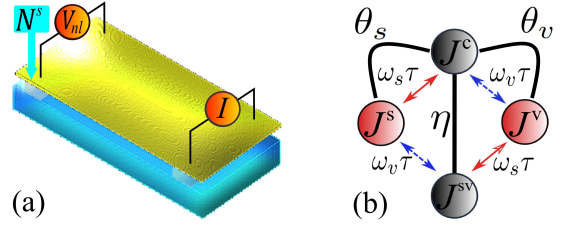}
\end{center}
\caption{(a) Sketch of a Hall-bar device used for measuring the nonlocal resistance $R_{nl}$: A current $I$ is injected on one side and a (non local) voltage  $V_{nl}$ is detected in the opposite side. The nonlocal resistance is defined as $R_{nl} \equiv V_{nl}/I$. In this work, we assume that  the spin and valley Hall effect coexist in the device. (b) Sketch of the four types of current response described by our model. Unlike the longitudinal electric (charge) current ($J^c$), the  transverse spin ($J^s$) and valley ($J^v$) currents and the longitudinal spin-valley ($J^{sv}$) current are all electrical neutral and therefore cannot be detected by all electrical means. In the absence of SHE (VHE), $\omega_v \tau$  ($\omega_s \tau$)
determines the conversion rate from $J^c$ to $J^v$ ($J^s$). However, when both SHE and VHE are present, $J^{sv}$ mediates a coupling between  $J^s$  and $J^v$, which has important consequences for the spin-diffusion as shown on the spin density of Fig. \ref{FIG2}(a).
}
\label{FIG1}
\end{figure}

Being electrically neutral, direct detection of spin and valley currents is not possible and their existence must be inferred by indirect means such as nonlocal transport measurements performed on a Hall bar device as depicted in Fig. \ref{FIG1}(a). In this setup, spin/valley currents are generated by driving an electric current between the two opposite right hand side contacts of the device. The neutral (spin/valley) currents  diffuse in the transverse direction to the applied electric current (field), leading to a charge accumulation and a nonlocal  voltage on the left hand side of the device. The nonlocal resistance (NLR) is defined as the ratio of the nonlocal voltage, $V_{nl}$ to the external current applied to the device, $I$.  Using this setup,  the VHE has been experimentally observed in devices made by depositing  monolayer graphene on hexagonal boron nitride (hBN)~\cite{gorbachev2014detecting}, bilayer graphene in a perpendicular displacement field \cite{shimazaki2015generation}, as well as optically pumped TMDs~\cite{mak2014valley,lee2016electrical,zeng2012valley,lee2017valley}. Likewise, the SHE has been experimentally observed in graphene decorated with absorbates~\cite{balakrishnan2014giant,balakrishnan_colossal,weeks2011engineering,ma2012strong} and graphene-TMDs heterostructures \cite{avsar2014spin,safeer2018room,benitez2018strongly}.

In connection to the observation of the SHE, the Hanle effect (HE), i.e. the modulation of the NLR  as a function of an in-plane magnetic field is considered to be the hallmark of  the existence of spin currents~\cite{balakrishnan2014giant,balakrishnan_colossal,PhysRevLett.119.136804,abanin2009nonlocal}. However, the absence of HE in some experiments in which a large enhancement of the NLR was observed~\cite{Tobias2018absence,kaverzin2015electron,neutral2015wang} hints at the existence of additional contributions to the NLR that are insensitive to the magnetic field.  One candidate that can contribute to the NLR is a valley current,  which, as we have shown elsewhere~\cite{zhang2016valley}, can arise from  a modest amount of nonuniform strain present in the Hall bar device. 

 Previous theoretical studies of nonlocal transport have focused  either on the VHE  \cite{beconcini2016nonlocal,zhang2016valley,song2018low} or the SHE \cite{abanin2009nonlocal,PhysRevLett.119.136804}. Building upon and largely extending earlier work, here we study the interplay between the two effects. In connection to the experiments described above, we show that this interplay can have nontrivial consequences for the spin transport in 2D materials. For instance, we find that spin density along the Hall bar can be modulated by the coupling between spin and valley currents, which can be  controlled by the application of  a nonuniform strain to the device~\cite{zhang2016valley}. 
This provides an  exciting link between spintronics and straintronics \cite{guinea2010energy,vozmediano2010gauge,amorim2016novel,Cazalilla2014qshe}.
In addition, 
we find that the HE may be strongly suppressed by the interplay with the VHE, and even absent under some circumstances. This finding can reconcile the apparently contradictory experimental results of various groups~\cite{kaverzin2015electron,avsar2015enhanced,neutral2015wang}, some of which have observed a large enhancement of the NLR but failed to observe the HE \cite{kaverzin2015electron,neutral2015wang}. Thus, the study reported here can be useful in guiding future studies of nonlocal transport in graphene, TMDs, and other 2D materials.

\textit{Theory:}  We shall work in  the diffusive regime  where $ k_F \ell\gg 1$, $k_F$ being the Fermi momentum of the electrons and $\ell$ the elastic mean-free path. This is the relevant regime to the devices that are experimentally studied (e.g. Refs.~\cite{balakrishnan2014giant,balakrishnan_colossal,kaverzin2015electron,avsar2015enhanced,neutral2015wang}). In this regime, the  transport of spin and valley degrees of freedom can be described by a set of diffusive equations. The latter can be derived microscopically from the Boltzmann equation\cite{huang2016direct,zhang2016valley,PhysRevLett.119.136804} or the Kubo formalism~\cite{PhysRevB.70.155308,PhysRevLett.105.066802}. In the steady state, the diffusion equations describing diffusion of spin and valley  take the following generic structure:
\begin{equation} \label{eq:eom_chargemt}
\mathcal{D}  \partial _{i}N^{\mu}- \sigma_D E_{i}^{\mu }=\left[
- \delta ^{\mu}_{\nu}\delta _{ij} +\left(R_{H}\right)^{\mu}_{\nu}\epsilon
_{ij}\right] J_{j}^{\nu}  .
\end{equation}
In the above set of equations, we have used the convention that repeated indices are summed over. 
The Latin indices correspond to the spatial component of the current, or field, i.e. $\{i,j\}\in \{x,y\}$ and $\epsilon _{ij}$ is the antisymmetric 2D Levi-Civita tensor. The Greek indices of the currents,  $\vec{J}^{\mu}$,  and densities, $N^{\mu}$, take values from the set $\{c,sv,v,s\}$. The latter stands for for charge ($c$), spin-valley ($sv$), valley ($v$), and spin ($s$) current (density) respectively. Note that the spin-valley current $\vec{J}^{sv}$ and density $N^{sv}$ must be included in the above hydrodynamic description as they can be excited when the spin splitting energy is much smaller than $\hbar/\tau$, where $\tau$ is the elastic scattering time.

The left hand side of Eq.~(\ref{eq:eom_chargemt}) contains the driving terms that result from spatial non-uniformity of the densities  $\propto \mathbf{\partial}_i N^{\mu}$ and  the generalized electric fields $\propto E_i^{\mu}$ (to describe real devices, we shall set $E_i^{\mu} =0$ for all $\mu \neq c$). The Drude conductivity $\sigma _{D}=ne^{2}\tau /m$ and the diffusion constant $\mathcal{D}=v_F^2 \tau/2$, which for the sake of simplicity we shall assume to be equal for all types of currents.  The right hand side of Eq.~(\ref{eq:eom_chargemt}) describes the effective Lorentz forces as well as current relaxation. We shall assume the relaxation rates for all currents are the same and equal to the Drude relaxation time $\tau$ (which is related to the mean-free path by $\ell = v_F \tau$ where $v_F$ is the Fermi velocity). This, together with the assumption of equal diffusion coefficients can be relaxed, and will not alter our conclusions qualitatively. Next, we introduce the coupling between different currents via the Hall resistivity matrix $R_{H}$ which describes both SHE and VHE. The latter couples the  charge ($c$, 1st row) and spin-valley currents ($sv$, 2nd row) to  valley ($v$, 3rd row) and spin ($s$, 4th row) currents:
\begin{equation}
R_{H}=%
\begin{bmatrix}
0 & 0 & \omega _{v}\tau & \omega _{s}\tau \\
0 & 0 & \omega _{s}\tau & \omega _{v}\tau \\
\omega _{v}\tau & \omega _{s}\tau & 0 & 0 \\
\omega _{s}\tau & \omega _{v}\tau & 0 & 0
\end{bmatrix}
\;\;
\begin{matrix}
c \\
sv \\
v\\
s
\end{matrix}
\end{equation}
The SHE  (VHE) can be regarded as emerging
from an effective spin (valley) dependent Lorentz force \cite{shen2005spin,chunli2016graphene,mak2014valley,zhang2016valley}. 
In $R_H$, the magnitude of such forces are parameterized by the ``cyclotron'' frequencies $\omega _{s}$ and $\omega_{v}$, for spin and valley, respectively.
These forces can have their origin in intrinsic or extrinsic SOC for the SHE \cite{sinova2015spin}, and in nonuniform strain~\cite{zhang2016valley} or skew scattering with  impurities in gapped (monolayer/bilayer) graphene (valley)~\cite{ando2015theory,ishizuka2017}. In the latter case, we neglect intrinsic Berry-curvature contributions to the valley current, as they are subdominant in the limit where impurities are dilute~\cite{ishizuka2017}. 
Note that when the valley and spin Hall effects coexist, the effective Lorentz force driving the VHE (SHE) current will act on the spin (valley) current. This is described by the additional entries in the $R_H$ which are not present when only the SHE or the VHE exist in the material (see Fig.~\ref{FIG1}(b)).

 In order to describe spin-valley transport with the above equations, we invert the resistivity matrix $R_{H}$ in the right hand side of Eq.~\eqref{eq:eom_chargemt}  and solve for the currents $J_{i}^{\mu}$:
\begin{equation} \label{eq:J_newmt}
J_{i}^{\mu }=-\left(D_{ij}\right)^{\mu}_{\nu }\partial _{j}N^{\nu}+\left(\sigma _{ij}\right)^{\mu}_{\nu
}E_{j}^{\nu }. 
\end{equation}
Note that the diffusion matrix is a rank-$2$ tensor in  the Latin indices $i,j$, and therefore it can be split into a symmetric ($\propto\delta _{ij}$) and antisymmetric ($\propto \epsilon_{ij}$) part according to $ D_{ij}=D_{0}\delta _{ij}+D_{H}\epsilon _{ij}$, where
\begin{equation}  \label{eq:D0mt}
D_{0}=\mathcal{D}_{r}%
\begin{bmatrix}
1 & \eta  & 0 & 0 \\
\eta  & 1 & 0 & 0 \\
0 & 0 & 1 & \eta  \\
0 & 0 & \eta & 1%
\end{bmatrix},
\end{equation}
\begin{equation} \label{eq:D1mt}
D_{H}=\mathcal{D}_{r}
\begin{bmatrix}
0 & 0 & \theta_{v} & \theta_{s} \\
0 & 0 & \theta_{s} & \theta_{v} \\
\theta_{v} & \theta_{s} & 0 & 0 \\
\theta_{s} & \theta_{v} & 0 & 0%
\end{bmatrix},
\end{equation}
\begin{equation}  \label{eq:D3mt}
\mathcal{D}_{r}=\mathcal{D}\frac{1+(\omega_{v}\tau )^{2}+(\omega_{s}\tau )^{2}}{[1+(\omega
_{v}\tau )^{2}+(\omega _{s}\tau )^{2}]^{2}-4\omega_{s}\omega_{v}\tau ^{2}}.
\end{equation}
Similarly, a decomposition of conductivity matrix as 
$\sigma_{ij} = \sigma_0 \delta_{ij} + \sigma_H \epsilon_{ij}$ 
can be obtained by replacing in the above expressions the diffusion constant $\mathcal{D}$ with the Drude conductivity $\sigma_{D}$. Note that the diffusion equation~\eqref{eq:J_newmt} involves an off-diagonal diffusion coefficient (cf. Eqs.~\ref{eq:D0mt} to \ref{eq:D3mt}) and  conductivity, which reduces to the well known limits. Thus, it yields the spin diffusion equations for a 2D electron gas \cite{raimondi2012su2,chunli2016graphene} when the second and third rows and columns of the diffusion matrix $D$ vanish. However, when the entries of the second and fourth rows and columns of $D$ vanish, Eq.~\eqref{eq:J_newmt} describes the diffusion of valley polarization.

In order to understand some of the important consequences of the coupling of spin and valley Hall effect, let us first solve Eq.~\eqref{eq:J_newmt} in the spatial uniform case where $\partial_j N^{\mu} = 0$. The ratios of the induced current (spin-valley $\vec{J}^{sv}$, valley $\vec{J}^{v}$, spin current  $\vec{J}^{s}$) over charge current $\vec{J}^{c}$ are the figures of merit for the various effects and they are denoted respectively as $\eta, \theta_s,\theta_v$; in particular, $\theta_s$ and $\theta_v$ are the spin Hall and valley Hall angles; $\eta$ describes the  conversion efficiency of the electric current to the spin-valley current, and it is given by the following expression:
\begin{equation} \label{SVC}
\eta =-\frac{2\,(\omega _{v}\tau) 
\,(\omega _{s}\tau) }{1+(\omega _{v}\tau
)^{2}+(\omega _{s}\tau )^{2}}.  
\end{equation}
Note that $\eta$ is proportional to the product of $\omega_v\tau$ and $\omega_c\tau$, meaning it is 
not zero provided that both SHE and VHE coexist.
As shown in Fig.~\ref{FIG1}(b), the generation of the spin-valley current is a two-stage process requiring the generation of a spin  (valley) current from driving electric  current via the SHE (VHE).  The resulting transverse current is then again deflected by the effective Lorentz force that causes the VHE (SHE) resulting in a \textit{longitudinal} spin-valley current. The factor of two in Eq.~\eqref{SVC} stems from the two possible routes by which this spin-valley conversion can take place: charge to spin to spin-valley and charge to valley to spin-valley (see Fig.~\ref{FIG1}(b)). 

 Furthermore, due to the spin-valley interplay, the valley ($\theta_{v}$) and spin Hall ($\theta_{s}$) angles are modified as follows:
\begin{align}
\theta _{v}&=\frac{1+(\omega _{v}\tau )^{2}-(\omega _{s}\tau )^{2}}{%
1+(\omega _{v}\tau )^{2}+(\omega _{s}\tau )^{2}}\,\omega _{v}\tau,
\label{VHE}\\
\theta _{s}&=\frac{1+(\omega _{s}\tau )^{2}-(\omega _{v}\tau )^{2}}{%
1+(\omega _{v}\tau )^{2}+(\omega _{s}\tau )^{2}}\,\omega _{s}\tau .
\label{SHE}
\end{align}
As expected, the spin (valley) Hall angle reduces to the familiar form $\theta_{s}=\omega _{s}\tau$ ($\theta_{v}=\omega _{v}\tau$) only when $\eta\propto \omega_s\omega_v=0$. However, in general $\theta_s$ ($\theta_v$)  deviates from their ``bare'' values due to the interplay of the spin and valley Hall effects. 
In typical spintronic materials, $\omega_{s}\tau \ll 1$ \cite{sinova2015spin}. However, nonuniform strain in graphene ~\cite{zhang2016valley}, for instance, can yield large values of the (bare) Hall angles for which $\left|\omega_v \tau \right| \sim 1$. 
In this case,  $\left| \theta_{s} \right| \sim \left| \omega_s \tau \right|^3 \ll 1$ implying that the  spin current will be strongly suppressed.

\begin{figure*}[t]
\begin{center}
\includegraphics[width=0.75\textwidth]{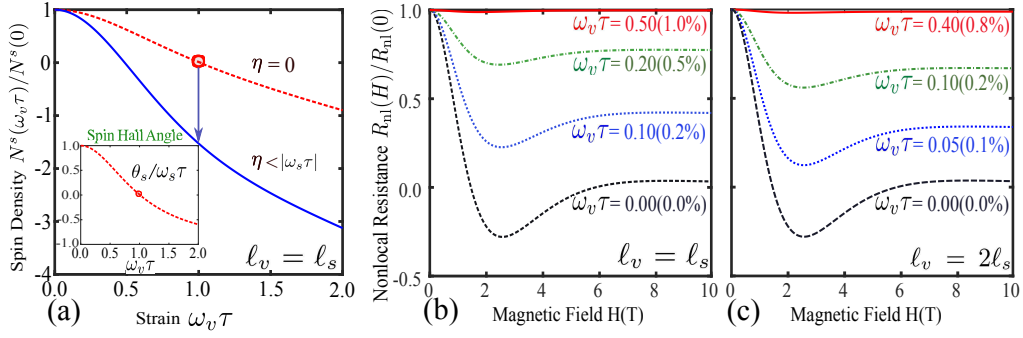}
\end{center}
\caption{(Color online) (a) Spin polarization, $N^{s}(x)$ at $x=2\, \mu$m for a Hall bar device of width $w = 0.5\, \mu$m versus $\omega_v \tau$, which is controlled
by the non-uniform strain applied to the device ($\tau$ is mean elastic collision time). 
 The red dotted line is plotted by artificially setting the coupling $\eta$ that controls the interplay of spin and valley to zero.  Notice that ignoring the interplay when solving the diffusion equations (cf. Eq.~\ref{2.5mt})  results in a substantial difference in the value of the spin-density diffusing along a Hall bar device.  The inset shows the spin Hall angle $\protect\theta_{s}$ from Eq.~\eqref{SHE} normalized to $\omega_{s}\tau \simeq -0.12$. Notice that a modest nonuniform strain can lead to a large valley Hall effect~\cite
{zhang2016valley}, $\omega _{v}\tau (\gtrsim 1)$. $\omega _{v}\tau =1$ can be induced by e.g. applying a nonuniform (uniaxial) strain of $2.0\%$ to a ribbon of width $w = 0.50 \mu$m. Panels (b) and (c) show the nonlocal resistance $R_{nl}(x,H)$ (normalized to 
 $R_{nl}(x,H=0)$) plotted versus the in plane magnetic field $H$ for two different values of the ratio of the valley to spin diffusion lengths: (b)
for $\ell _{v}=\ell _{s}$ and (c) for $\ell _{v}=2\ell _{s}$. Parameters: $\ell _{s}=0.53\,\mu $m, $x=2.00\,\mu$m and $y=0.25
\, \mu$m. }
\label{FIG2}
\end{figure*}

\textit{Control of spin diffusion by means of strain:} Next, we study the consequences of the interplay between spin and valley Hall effects for the spin transport. We first derive the drift-diffusion equations by supplementing Eq.~\eqref{eq:eom_chargemt} with the steady-state
continuity equations for the currents, i.e.  $\partial_{i}J_{i}^{\mu}=-\delta ^{\mu}_{\nu }/\tau ^{\nu }N^{\nu}$, where the limit $\tau^{\mu}\to +\infty$ for $\mu = c$  must be taken since the electric current is strictly conserved. Hence,
\begin{equation} \label{eq:ddemt}
\left[ \left(D_{0}\right)^{\mu}_{\nu }\nabla^{2}-\frac{\delta ^{\mu}_{\nu }}{
\tau ^{\nu }} \right] N^{\nu} = S^{\mu}.
\end{equation}
In the above equations, $\tau^{\mu}$ are the relaxation times of the various currents. We have also assumed that spin-charge 
conversion mechanisms like the Edelstein effect or the direct magneto-electric coupling~\cite{raimondi2012su2,PhysRevLett.119.136804}  can be neglected in a first
approximation. $S^{\mu}$ is a source term given by
\begin{equation}
S^{\mu} = \epsilon _{ij} \left[ -\partial _{i}\left(D_{H}\right)^{\mu}_{\nu}\partial
_{j}N^{\nu}+\partial _{i}\left(\sigma _{H}\right)^{\mu}_{\nu} E_{j}^{\nu}\right].	
\end{equation}
Note that $S^{\mu}$ vanishes in the bulk of the Hall bar device, and it is only nonzero wherever $D_H$ and $\sigma_H$ are discontinuous, i.e. at the boundary.  Thus, away from the boundaries, $D_H$ and $\sigma_H$ become homogeneous and Eq. \eqref{eq:ddemt}  can be written as follows:
\begin{equation} \label{2.5mt}
\nabla^2 N^{\mu }-\mathcal{M}^{\mu}_{\nu }N^{\nu }=0,  
\end{equation}
where
\begin{equation}  \label{2.6mt}
\mathcal{M}^{\mu}_{\nu}=\frac{1}{1-\eta ^{2}}%
\begin{bmatrix}
\ell _{v}^{-2} & -\eta\ell _{s}^{-2} \\
-\eta\ell _{v}^{-2} & \ell _{s}^{-2}
\end{bmatrix}
\;\;
\begin{matrix}
v \\
s
\end{matrix} .
\end{equation}
Only spin and valley densities are considered in the above equations  because they are the only responses in the transverse direction to the applied electric field. In this expression, $\ell _{v}=\sqrt{\mathcal{D}
_r\tau ^{v}}$ ($\ell _{s}=\sqrt{\mathcal{D}_{r}\tau ^{s}}$) is valley (spin) relaxation length and $\mathcal{D}_r$ is the (renormalized) diffusion constant (cf. Eq.~\ref{eq:D3mt}). Note that the off-diagonal term $\eta$ mixes the spin and valley densities. Eq.~\eqref{2.5mt} are solved by diagonalizing the diffusion matrix, such that 
$\mathcal{M}^{\mu}_{\nu}|\vec{\hat{e}}^{\nu}_{a}\rangle =\mathcal{L}
_{a}^{-2}|\vec{\hat{e}}^{\mu}_{a}\rangle $, where $\mathcal{L}_a$ ($a=1,2$) corresponds to the  diffusion length of the eigenmode $|\vec{\hat{e}}^{\mu}_{a}\rangle$.

 In order to illustrate the properties of the solution to the above diffusion equations, we consider a non-uniformly strained graphene device decorated with absorbates that locally induce SOC. As mentioned above, this system can be relevant to the experiments reported  in Refs.~\cite{neutral2015wang,kaverzin2015electron}.  In the long wave-length limit, the effect of nonuniform strain
can be described by a out-of-plane (orbital) pseudo-magnetic field, which takes opposite signs at opposite valleys~\cite{guinea2010energy,vozmediano2010gauge,zhang2016valley}. In earlier work, we have shown that modest amounts of nonuniform strain can lead to a sizable VHE~\cite{zhang2016valley}. In addition, skew 
scattering with the absorbates induces the
SHE~\cite{ferreira2014extrinsic,balakrishnan2014giant,yang2016extrinsic,chunli2016graphene}. Thus, in this system both VHE and SHE  coexist and the spin and valley transport is described by Eq. \eqref{2.5mt}, whose solution we shall analyze in what follows.  

For the sake the simplicity, we take  the Hall bar to be an infinitely long 
conducting channel of width $w$~\cite{abanin2009nonlocal,beconcini2016nonlocal,SM}. The solution 
of the coupled diffusion equations is  simplified by setting $N^{c}(\vec{r}) =0$, which results from assuming  the complete screening of the electric field inside the  metal. Thus, the electrostatic potential $\Phi \left( \vec{r}\right) $ obeys the Laplace
equation, i.e. $\nabla^{2}\Phi \left( \vec{r}\right) =0$. Using the appropriate boundary conditions \cite{beconcini2016nonlocal,zhang2016valley}, the valley and spin densities, at the edge ($y=\pm w/2$), are given by the following 
expression:
\begin{align}
N^{\mu}(x)&=\frac{iI}{\mathcal{D}_{r}} \sum_{\nu,b} \vec{\hat{e}}^{\mu}_{b} (\vec{\hat{e}}^{-1})_{\nu}^{b}
\theta _{\nu}\int_{-\infty }^{\infty}\frac{dk  }{2\pi k } 
\frac{e^{ikx} \mathcal{F}_b\left( k\right)}
{1+\sum\limits_{a}\Theta _{a}^{2}\mathcal{F}_a\left( k\right) } , \notag\\
\mathcal{F}_a( k)&=\frac{k\tanh ( kw/2)}{\kappa
_{a}\tanh ( \kappa _{a}w/2)},
\label{SVD} 
\end{align}
with $a,b=(1,2)$. $\mu=v,s$ correspond to valley and spin densities, respectively. $\kappa _{a}=\sqrt{k^{2}+\mathcal{L}
_{a}^{-2}}$ and $\Theta _{a}^{2} = \left[
\vec{\hat{e}}^{\mu}_{a} \theta_{\mu}  \right] [ (\vec{\hat{e}}^{-1})_{a}^{\nu}
\theta_{\nu}]$.

Using the above results, we show in what follows that  the spin polarization diffusing in the Hall bar   
can be controlled by the application of  nonuniform strain. In Fig.~\ref{FIG2}(a), we plot the 
spin polarization  $N^{s}(x,y)$ (taking $x = 2\, \mu$m and $y = 0.25\, \mu$m)
as a function of the $\omega_v \tau$, which is determined by the strength of the pseudo-magnetic field  induced by the nonuniform strength applied to device~\cite{zhang2016valley}. 
Interestingly, the spin polarization does not vanish even when the strain is tuned to make the effective spin Hall angle  (cf. Eq.~\eqref{SHE}) $\theta_s = 0$ (see red circle in the inset of Fig.~\ref{FIG2}(a)). This is a dramatic consequence of the coupling between the SHE and VHE, whose strength is measured by $\eta$ (cf. Eq. \eqref{SVC}).
Due to this coupling, the valley density accumulation induced by VHE can be converted to spin density. Note that if we  solve the diffusion equations by ignoring the spin-valley coupling (i.e. by artificially setting the parameter $\eta =0$) the behavior of the spin polarization (red line in Fig. \ref{FIG2}(a)) would be very different.

\textit{Suppression of the Hanle effect:} 
Finally, we show that the interplay between the SHE and VHE  can lead to the suppression of the HE.  As mentioned above, the quantity of experimental interest is  the NLR of the Hall bar measured at distance $x$ from the current injection point. The HE results in the appearance of an oscillatory component in the NLR as a function of the external magnetic field $H$ applied in the plane of the device. The oscillation is the result of the  precession of the electron spins in the external magnetic field $H$. 

 By solving the coupled diffusion and Laplace equations, the NLR can be obtained from: 
\begin{equation}
R_{nl}\left( x, H\right) = \frac{1}{I}\left[\Phi\left( x,-\frac{w}{2},H\right) -\Phi \left( x,\frac{w}{2},H\right)\right]
\end{equation}
where $\Phi(x,y,H)$ is the electrostatic potential for 
an in-plane magnetic field $H$. In the absence of both SHE and VHE, the NLR is given by the van der Paw law: $R_{nl}^{0} \simeq \frac{4}{\pi \sigma^{c}}  e^{-\left\vert x\right\vert / \mathcal{L}_{0}},$ for $|x| \gg \mathcal{L}_{0}=w/\pi $. However, experimentally it is found~\cite{balakrishnan2014giant,balakrishnan_colossal,PhysRevLett.119.136804,kaverzin2015electron} that the NLR is greatly enhanced with respect to the \emph{Ohmic} signal. When the spin-diffusion length 
$\ell_s$ is  shorter than the valley-diffusion length, 
$\ell_v$, a suppression of the HE is   expected. This is because the valley currents, which diffuse much farther and therefore will
yield the dominant contribution to $R_{nl}(x,H)$, are completely insensitive to the in-plane magnetic field. Strikingly, we find  that for $\ell_{v}$ and $\ell_s$ take comparable values, the HE can be suppressed by a moderate amount of nonuniform strain present in the device. 

In order to compute $R_{nl}(x,H)$  we add to  the diffusion equations~\eqref{eq:ddemt}, a Zeeman term, which induces precession. A sufficiently strong magnetic field $\vec{H}\propto \mathbf{\hat{y}}$ converts the out-of-plane spin polarization, $N^s(x)$, into an in-plane spin polarization (along the $x$-direction). Since the nonlocal voltage is determined by the magnitude $N^s(x)$ at the location the voltage probes (see \cite{SM}), this results in the NLR developing an oscillatory component. When the SHE and VHE coexist, describing precession requires that we account for the diffusion of the components of the spin  and spin-valley densities in the plane perpendicular to  $\vec{H}$. The solution of the resulting diffusion equations becomes more involved and the details are provided in~\cite{SM}. Here we focus on the discussion of the result for the NLR, which is shown in Fig.~\ref{FIG2}(b,c).

 In Fig.~\ref{FIG2}(b), the NLR versus the applied magnetic field $H$ has been plotted for 
$\ell_v=\ell_s$. Setting $\omega_{v}\tau =0$, we  recover the result obtained by Abanin et al.~\cite{abanin2009nonlocal}, showing the characteristic oscillatory component in $R_{nl}(x,H)$ associated with the HE. By applying  an increasing amount of nonuniform strain to the Hall bar (i.e. increasing $\omega_v\tau$),  the amplitude of the oscillatory component in the NLR is suppressed and almost disappears for $\omega_v \tau \sim 0.5$, which, for typical experimental parameters~\cite{kaverzin2015electron}, corresponds to a nonuniform (uniaxial) strain of $\approx 1\%$ applied to a Hall bar $0.5\: \mu$m wide.  Thus, the suppression of the HE happens due to the competition between the spin and valley Hall effects. As mentioned above, when $\omega_v\tau \sim  1$, the  spin Hall angle $\theta_s$ (cf. Eq.~\ref{SHE}) is  strongly reduced, see Eq.~\eqref{SHE}. Since the magnitude of $\theta_s$ determines the HE, the existence of a sizable VHE resulting from strain can suppress the HE. For larger valley diffusion length ($\ell_v=2\ell_s$), the suppression of the HE becomes even more obvious and happens for smaller amount of strain, as shown in Fig.~\ref{FIG2}(c). In~\cite{SM} we show that the suppression of the HE is not affected by charging the carrier density or sign. Notice that the moderate amounts of strain considered here could be unintentionally introduced during the process of device fabrication.  
Thus, our findings are relevant for the interpretation of some of the nonlocal transport measurements in graphene
decorated with hydrogen \cite{neutral2015wang} and gold adatoms 
\cite{kaverzin2015electron}, where a large enhancement of the NLR  was detected without HE. 

 Before concluding, it is worth commenting on other possible causes for the suppression of the HE. Indeed,  suppression of the effect may also arise from  a sizable spin-valley locking such as the one present in the band structure of TMDs~\cite{suzuki2014valley}. Effectively, this type of spin-valley locking can be described as a Zeeman coupling to an out-of-plane magnetic field which takes opposite signs at opposite valleys.  However, in graphene devices, such type of spin-valley would require  breaking  the sublattice symmetry, which can be induced by either the substrate or the absorbates decorating the device. However, such a strong sublattice symmetry breaking was not
experimentally observed~\cite{kaverzin2015electron}.

\textit{Summary and outlook-} We have explored  a number of important  consequences of the coexistence of spin and valley Hall effects in a two-dimensional material:  We have shown the latter leads to  the emergence of neutral longitudinal spin and valley polarized current. Furthermore, we have shown the spin polarization diffusing in the material can be controlled by means of nonuniform strain. Finally, we have shown the Hanle effect in response to an in-plane magnetic field can be strongly suppressed due to the competition of the two effects. We believe the suppression  of the Hanle effect noticed here  will shed light on experimental controversies concerning the origin of the enhancement of the nonlocal resistance in various types of graphene  devices 
~\cite{balakrishnan_colossal,balakrishnan2014giant,kaverzin2015electron,avsar2015enhanced,neutral2015wang}. The theory presented here can also be extended in various other directions, such as accounting for other  spin-charge conversion mechanisms beyond the SHE (such as the inverse spin-galvanic effect) and a weak spin-valley (Zeeman) coupling, which is present in hybrid graphene-TMD structures. Both effects are  expected to be important when spatial inversion symmetry is broken.

\emph{Acknowledgments-}This work is supported by the Ministry of Science and Technology (Taiwan) under contract number NSC 102- 2112-M-007-024-MY5 (MAC and CH), the Spanish Ministerio de Economia y Competitividad (MINECO) through Project No. FIS2014-55987-P and FIS2017-82804-P (XP and CL), and Taiwan's National Center of Theoretical Sciences (MAC and CL). We thank F. Guinea, A. Kaverzin, and R. Stephen for useful discussions.
%


\section{Supplement Materials of Control of Spin Diffusion and Suppression of the Hanle Effect by the Coexistence of Spin and Valley Hall Effects} 
\setcounter{section}{0}
\setcounter{equation}{0}
\renewcommand{\theequation}{A.\arabic{equation}}

\section{Kinetic theory}
\subsection{Boltzmann equation} \label{sec:BE} 
In this subsection, we introduce a quantum Boltzmann equation (QBE) capable of describing a system in which both spin (SHE) and valley Hall (VHE) effects co-exist:
\begin{equation} \label{QBE00}
 \dot{n}_{\vec{k}}+%
\vec{v}^{\vec{k}}\cdot \nabla_{\vec{r}}  n_{\vec{k}} +\vec{F}_{\vec{k}}\cdot 
\mathbf{\nabla}_{\vec{k}}n_{\vec{k}} +i\omega_L\left[ n_{\vec{k}},\vec{s}\cdot \vec{m}\right] = \mathcal{I}_r\left[n_{\vec{k}}\right].
\end{equation}
In the above expression, the function $n_{\vec{k}}$ is the density-matrix distribution function of the carriers (electrons or holes) in the Bloch state characterized by (crystal) momentum $\vec{k}$. Thus, it is a $4\times 4$ matrix in spin-valley space. The force $\vec{F}$ driving the carrier motion can be split into three terms:
\begin{equation}
  \vec{F}_{\vec{k}}=\vec{F}^l_{\vec{k}}+\vec{F}^s_{\vec{k}}+\vec{F}^v_{\vec{k}},
\end{equation}
where 
\begin{align}
\vec{F}^l_{\vec{k}} &=\vec{F}^E_{\vec{k}}+\vec{F}^B_{\vec{k}}\equiv e\vec{E}+e \vec{v}^{\vec{k}}\times \vec{B}, \label{eq:fl} \\
\vec{F}^s_{\vec{k}} &= e \vec{v}^{\vec{k}}\times (\hat{s}_{z}\vec{\mathcal{B}}_s), \label{eq:fls} \\
\vec{F}^v_{\vec{k}} &=e  \vec{v}^{\vec{k}}\times (\hat{\tau}_{z}\vec{\mathcal{B}}_v).   
\end{align}
The $F^l_{\vec{k}}$ is the electromagnetic Lorentz force due to  external (in-plane)
electric and (out-of-plane) magnetic fields ($\vec{E}\perp \vec{\hat{z}}$ and $\vec{B}\parallel \vec{\hat{z}}$, respectively). While $F^{s/v}_{\vec{k}}$ are the effective (Lorentz-like) forces for effective (out-of-plane) spin/valley magnetic field ($\vec{\mathcal{B}}_s\parallel \vec{\hat{z}}$ and $\vec{\mathcal{B}}_v\parallel \vec{\hat{z}}$, respectively), from which the SHE and VHE originate. In Eq.~\eqref{eq:fl}, $e(<0)$ is the charge of the electron, $\vec{v}^{\vec{k}}  = \hbar^{-1} \nabla_{\vec{k}}\epsilon_{\vec{k}}$ is the velocity of electron with  (crystal) momentum $\vec{k}$, and $\epsilon_{\vec{k}}$ is the band dispersion. We assume that there is  no Berry curvature in the band and therefore  anomalous velocity vanishes. $\hat{s}_{a}$, $\hat{\tau}_{a}$ $\left(
a=o,x,y,z\right) $ are Pauli matrices describing the spin and valley (pseudo-spin), respectively. The matrix $s_o$ ($\tau_o$) corresponds to the spin (valley) unit matrix.

The magnitude of the SHE (VHE) has been parameterized in the above equations by the effective spin (valley) magnetic field $\hat{s}_{z}\vec{\mathcal{B}}_s$ ($\hat{\tau}_{z}\vec{\mathcal{B}}_v$), which points in opposite directions for electrons of different spins (valleys). The last term of the left hand side of Eq. (\ref{QBE00}) describes  spin precession with a Larmor frequency  $\omega _{L}= \textsl{g}\mu_B H /\hbar$, which is proportional to the magnitude of the total applied (Zeeman) magnetic field $\vec{H}$ ($\textsl{g}$ is the gyromagnetic factor and $\mu_B$ is the Bohr magneton). 
In Eq.~\eqref{QBE00}, $\vec{m} = \vec{H}/H$ denotes the direction of the total magnetic field and $\vec{B}$ in
Eq.~\eqref{eq:fl} denotes the component of the magnetic field perpendicular to the plane of the material. 
In what follows, we shall assume that the external magnetic field (when present) is applied in the plane of the 2D system, which means $\vec{B} = 0$ and therefore the magnetic field part of Lorentz force $\vec{F}^B_{\vec{k}} = 0$.

On the right hand side of Eq. \eqref{QBE00} $\mathcal{I}_r\left[n_{\vec{k}}\right]$ is the  (dissipative) collision integral. Strictly speaking, the force terms proportional to $\vec{F}^{s/v}_{\vec{k}}$  can arise from the collision integral as a result of skew scattering  (see Sec.~\ref{sec:ex} below and e.g. Refs.~\cite{chunli2016graphene,zhang2016valley}). Alternatively, a weak uniform (i.e. intrinsic) Rasbha-type SOC can also give rise to a Lorentz-like force term like $\vec{F}^{s}_{\vec{k}}$ in the QBE~\cite{raimondi2012su2,huang2017spin}. Furthermore, nonuniform strain can give rise to a force like $\vec{F}^{v}_{\vec{k}}$ (see below, Sec.~\ref{sec:ex}, and \cite{zhang2016valley}). 

\subsection{Linearized Boltzmann equation}
For small applied electric field, $\vec{E}$, the solution to the QBE~\eqref{QBE00}, can be
obtained by using the following ansatz for electron density-matrix distribution function:
\begin{equation}
 n_{\vec{k}}(\vec{r},t)=n^{0}\left[ \epsilon _{k}-\mu_F
- \gamma_{\nu}(\mu ^{\nu}\left( \vec{r}%
,t\right) +\vec{v}^{\nu}\left( \vec{r},t\right) \cdot \vec{k)}\right].
\end{equation}
In the above equation, 
$n^0(\epsilon)$ is Fermi-Dirac distribution at the absolute temperature $T$
and global chemical potential $\mu_F$. 
The convention of 
summing over repeated Greek indices like $\nu$ has been used,   with matrix $\gamma_{\nu}$ belonging to the set of $4\times 4$ matrices $\{\hat{s}_{o},\hat{s}_{z}\}\otimes \{%
\hat{\tau}_{o},\hat{\tau}_{z}\}=\{\hat{s}_{o}\hat{\tau}_{o},\hat{s}_{o}\hat{%
\tau}_{z},\hat{s}_{z}\hat{\tau}_{o},\hat{s}_{z}\hat{\tau}_{z}\}$, which are a set
of $4\times 4$ matrices in spin-valley space. The index $\nu$ runs over the combinations for charge ($c = oo$), spin-valley ($sv=zz$), valley ($v=oz$) and spin ($s=zo$) indices.
 The fields
$\vec{v}^{\nu}(\vec{r},t)$ and $ \mu^{\nu}\left( \vec{%
r},t\right) $ correspond to the drift velocity of the electron fluid and the local chemical potential,
respectively. Both are proportional to applied electric field, i.e., $\vert\vec{v}^{\nu}(\vec{r},t)\vert \propto \vert\vec{E}\vert$ and $ \mu ^{\nu}\left( \vec{%
r},t\right)\propto |\vec{E}|$.  To linear order in $\vec{v}^{\nu}(\vec{r},t)$ and $ \mu^{\nu}\left( \vec{%
r},t\right) $, the
deviation of distribution function from its equilibrium, $\delta n_{\vec{k}
}= n_{\vec{k}}-n_{k}^{0}$ reads:
\begin{equation} \label{Ans}
 \delta n_{\vec{k}}(\vec{r},t)\simeq
\gamma_{\nu}\left[ \mu ^{\nu}\left( \vec{r}%
,t\right) +\vec{v}^{\nu}\left( \vec{r},t\right) \cdot \vec{k}\right] %
\left[ -\partial _{\epsilon }n_{\epsilon }^{0}\right] _{\epsilon =\mu_F }, 
\end{equation}
with $\nu=(c,sv,v,s)$.  Hence, 
\begin{equation}
 \nabla_{\vec{k}} \delta n_{\vec{k}}(\vec{r},t)\simeq  
\gamma_{\nu}\vec{v}^{\nu}\left( \vec{r},t\right) 
\left[ -\partial _{\epsilon }n_{\epsilon }^{0}\right] _{\epsilon =\mu_F }.
\end{equation}
Thus, to linear order in $\vec{E}$, linearization of QBE yields:
\begin{widetext}
\begin{equation} \label{QBE0}
 \delta \dot{n}_{\vec{k}}+
\vec{v}^{\vec{k}}\cdot  \nabla_{\vec{r}} \delta n_{\vec{k}} +e\left(\vec{E}+\vec{v}^{\vec{k}}\times \vec{B}\right)\cdot 
\mathbf{\nabla}_{\vec{k}} n^0_{\vec{k}} +\vec{F}^s_{\vec{k}}\cdot 
\mathbf{\nabla}_{\vec{k}}\delta n_{\vec{k}}+\vec{F}^v_{\vec{k}}\cdot 
\mathbf{\nabla}_{\vec{k}}\delta n_{\vec{k}}+i\omega_L\left[ \delta n_{\vec{k}},\vec{s}\cdot \vec{m}\right] = \mathcal{I}_r\left[\delta n_{\vec{k}}\right],
\end{equation}
\end{widetext}
where we have used:
\begin{align}
 \vec{F}^s_{\vec{k}}\cdot 
\mathbf{\nabla}_{\vec{k}} n^0_{\vec{k}}\propto (\hat{\vec{k}}\times \hat{z} ) \cdot \hat{\vec{k}}&=0, \\
\vec{F}^v_{k}\cdot 
\mathbf{\nabla}_{\vec{k}} n^0_{\vec{k}}\propto (\hat{\vec{k}}\times \hat{z} ) \cdot \hat{\vec{k}}&=0,
\end{align}
together with the vanishing of the collision integral for the equilibrium distribution $n^0_{k}$.

\subsection{Example of a microscopic model}\label{sec:ex}
The above linearized QBE can be obtained for various types of microscopic models. 
In this subsection, we study an instance of much experimental interest describing a monolayer of graphene subject to nonuniform strain and decorated with adatoms. The latter induce spin-orbit coupling (SOC) by proximity to the graphene layer. For the sake of simplicity, the spatial dependence of SOC is approximated by a Dirac delta potential (but more complicated dependence will not alter our results qualitatively~\cite{ferreira2014extrinsic}). The spin-dependence corresponds to the so-called Kane-Mele SOC, which is known to lead to  \emph{extrinsic} SHE~\cite{chunli2016graphene,ferreira2014extrinsic}. 
\subsubsection{Pseudo-magnetic field in strained graphene}
Within  the $\vec{k}\cdot\vec{p}$ approximation to the band structure of graphene~(see e.g. \cite{katsnelson2012graphene}), nonuniform (shear) strain can be described as a pseudo-gauge field which takes opposite signs at opposite valleys~ (see  e.g. \cite{guinea2010energy,vozmediano2010gauge,katsnelson2012graphene,amorim2016novel}):
\begin{align}
 H_0[\vec{k}-\hat{\tau}_{z}\vec{\mathcal{A}}(\vec{r}) ]&=\hbar v_F \left[\hat{\tau}_{z} \hat{\sigma}_x (k_x-\hat{\tau}_{z}\mathcal{A}_x) \right. \notag\\
&\qquad\left.  + \hat{\sigma}_y (k_y-\hat{\tau}_{z}\mathcal{A}_y)\right].
\end{align}
In the above expression $v_F$ is the Fermi velocity and $\sigma_x,\sigma_y$ are the Pauli matrices describing the sublattice pseudo-spin.   The pseudo-gauge $\vec{\mathcal{A}}(\vec{r})$ field  which describes the (strain-induced) local displacement of the Dirac points at the two valleys is given by the following expression:
\begin{equation}
   \vec{\mathcal{A}}(\vec{r})=(\mathcal{A}_x,\mathcal{A}_y)= \tfrac{\beta}{ae}\left(u_{xx}-u_{yy},-2 u_{xy} \right),
\end{equation}
where $\beta=\tfrac{d \log t}{d \log a} \simeq 2
$, $t$ being the nearest neighbor hopping  amplitude, $a$ is the carbon-carbon distance, and 
\begin{equation}
    u_{ij}=\frac{1}{2}(\partial_i u_j + \partial_j u_i),
   \end{equation}
is strain tensor. 
Note that, since $u_{ij}$ is invariant (i.e. even) under time-reversal (TR) and $\hat{\tau}_{z}\vec{\mathcal{A}}(\vec{r})$ even under TR (recall that $\tau_z \to -\tau_z$ under TR). This is different from a real magnetic field, for which the gauge field is odd under TR. 

The pseudo-magnetic field that determines the valley Lorentz-like force, $\vec{F}^v_{\vec{k}}$ can be obtained from the standard expression:
\begin{equation} \label{vefB}
    \hat{\tau}_{z}\vec{\mathcal{B}}_v=\hat{\tau}_{z}\nabla \times \vec{\mathcal{A}}(\vec{r})=\hat{\tau}_{z}(\partial_x \mathcal{A}_y- \partial_y \mathcal{A}_x)\: \vec{\hat{z}}. 
\end{equation}
Thus, as mentioned above, the pseudo-magnetic field $\hat{\tau}_{z}\vec{\mathcal{B}}_v$ induced by nonuniform strain has opposite signs at opposite valleys as required by the fact that strain does not break TR invariance. In what follows, for the sake of simplicity, we shall assume that the pseudo-magnetic field $\hat{\tau}_{z}\vec{\mathcal{B}}_v$ is  spatially uniform, which requires  particular configurations of nonuniform strain~\cite{guinea2010energy,amorim2016novel,zhang2016valley}. Thus, we have shown how strain can give rise to an effective Lorentz-like force $\vec{F}^v_{\vec{k}}$, which drives the VHE (alternatively, this force can emerge from skew scattering with scalar impurities in bands with nonzero Berry curvature~\cite{ishizuka2017}). Via the semi-classical equations of motion~\cite{zhang2014direct}, the latter will enter the QBE in~\eqref{QBE00}.

\subsubsection{Adatom-induced SHE}

The spin transport  properties of graphene can be modified by the presence of adatom impurities~\cite{weeks2011engineering,ferreira2014extrinsic,yang2016extrinsic,chunli2016graphene}. 
In the dilute impurity limit, the dominant mechanism for the spin-charge conversion via the extrinsic SHE is skew scattering~\cite{Nagaosa_RevModPhys.82.1539}, which  effectively gives rise to a spin-dependent Lorentz-like force~\cite{chunli2016graphene}. 

 Within the $\vec{k}\cdot\vec{p}$ theory,
the potential for a single-impurity takes the following form: 
\begin{equation}
V(\vec{r})=(\mathcal{V}_{c}\hat{s}_{o}\hat{\tau}_{o}\hat{\sigma}
_{o}+\mathcal{V}_{s}\hat{s}_{z}\hat{\tau}_{z}\hat{\sigma}
_{z})R^{2}\delta \left( \vec{r}\right).
 \label{H1}
\end{equation}
In the above expression, $R$  is a length scale  of the order of the impurity radius. We shall assume that $R\gg a$, that is, much larger than the inter-carbon separation so that inter-valley scattering can be safely neglected~\cite{basko2008resonant} but
$R \lesssim 10$ nm, so that the potential 
can be approximated by a Dirac $\delta$-function. This approximation should be a good description of a monolayer of graphene decorated by adatom
clusters~\cite{ferreira2014extrinsic,balakrishnan_colossal}.  Hence,
upon solving the scattering problem, the  on-shell $T$-matrix projected on the carrier band can be obtained and reads:
\begin{align}
 T_{\vec{k}\vec{p}}^{+}=t_c(k)\hat{s}%
_{o}\cos(\theta_{\vec{k}\vec{p}}/2)+t_s(k)\hat{s}_{z}\sin(\theta_{%
\vec{k}\vec{p}}/2).
\end{align}
The functions $t_c(k)$ and $t_s(k) $ depend on momentum $k$ of the incoming electron and the impurity potential
parameters, i.e. $\mathcal{V}_c,\mathcal{V}_s$, in our model. See e.g. Refs.\cite{huang2016direct} and \cite{zhang2016valley} for the detailed expressions of these functions.

The effect of impurities is described by the
collision integral $\mathcal{I}\left[ \delta n_{\vec{k}}\right]$. The  complete form of the latter (which includes the dissipative $\mathcal{I}_r[\delta n_{\vec{k}}]$ introduced in Eq.~\ref{QBE00}) has been derived in Ref.~\cite{huang2016direct}, extending earlier work of Kohn and Luttinger in order to account for the effects of disorder on the electron internal degrees of freedom such as spin and valley pseudo-spin. 
To leading order in the density of impurities, $n_{\mathrm{i}}$, the  collision integral reads:
\begin{align}  \label{E.222}
\mathcal{I}[\delta n_{\vec{k}}]& =\frac{2\pi}{\hbar} n_{\mathrm{i}}\sum_{\vec{p}}\delta
(\epsilon _{k}-\epsilon _{p})\left[ T_{\vec{k}\vec{p}}^{+}\delta n_{\vec{p}%
}T_{\vec{p}\vec{k}}^{-}\right.   \\
& \left. -\frac{1}{2}\left\{ \delta n_{\vec{k}}T_{\vec{k}\vec{p}}^{+}T_{\vec{%
p}\vec{k}}^{-}+T_{\vec{k}\vec{p}}^{+}T_{\vec{p}\vec{k}}^{-}\delta n_{\vec{k}%
}\right\} \right] , \notag
\end{align}%
which is determined by the scattering data of a single scatterer.
Using the above
ansatz, Eq.~\eqref{Ans}, the collision integral (\ref{E.222}) reduces to: 
\begin{align} \label{3.28}
\mathcal{I}\left[ \delta n_{\vec{k}}\right] &=&\frac{\pi }{\hbar }n_{\mathrm{i}}\sum_{\vec{p}}\delta \left( \epsilon_p-\epsilon _
k \right) 2\hat{T}^+_{\vec{k}\vec{p}}\hat{T}^-_{\vec{k}\vec{p}}(\delta n_{\vec{p}}-\delta n_{\vec{k}}) , 
\end{align}
where
\begin{align} \label{3.250}
2\hat{T}_{\vec{k}\vec{p}}^{+}\hat{T}_{\vec{k}\vec{p}}^{- } &=\left[
\left\vert t_{c}\right\vert ^{2}\left( 1+\cos \theta \right)
+\left\vert t_{s}\right\vert ^{2}\left( 1-\cos \theta \right) %
\right] \gamma ^{c} \notag \\
&+2\text{\textrm{Im}}\left( t_{c}t_{s}^{\ast }\right)
\sin \theta \gamma ^{s}  
\end{align}
\begin{equation} \label{3.260}
\delta n_{\vec{p}}-\delta n_{\vec{k}}=\left[ -\partial _{\epsilon
}n^{0}\left( \epsilon \right) \right]_{\epsilon =\mu_F }\sum_{\nu}\gamma ^{\nu}\vec{v}^{\nu}\left( \vec{r}\right) \cdot \hbar (%
\vec{p}-\vec{k}).
\end{equation}
Substituting Eqs. \eqref{3.250} and \eqref{3.260} into   Eq. \eqref{E.222}, the collision integral takes the following form:
\begin{align}
\mathcal{I}\left[ \delta n_{\vec{k}}\right] &=\left[ \partial
_{\epsilon }n^{0}\left( \epsilon \right) \right] _{\epsilon =\mu_F }\hbar \vec{%
k}\cdot \left\{ \frac{\gamma ^{c}}{\tau}\sum_{%
\nu}\gamma ^{\nu}\vec{v}^{\nu}\left( \vec{r}\right)
\right.  \notag \\
&+\left.\gamma ^{s}\omega_{s} \sum_{\nu}\gamma
^{\nu}\vec{v}^{\nu}\left( \vec{r}\right) \times \hat{\vec{z}}\right\},
\end{align}
where 
\begin{align}
\frac{1}{\tau (k)}&=\frac{k n_{\mathrm{i}}}{4\hbar ^{2}v_{F}}\left[
\left\vert t_{c}(k)\right\vert ^{2}+3\left\vert t_{s}(k)\right\vert ^{2}\right] ,\\
\omega_{s}(k)&=\frac{k n_{\mathrm{i}}}{4\hbar ^{2}v_{F}}\left[-2\mathrm{Im}
 \{t_{c}(k) t_{s}(k)\}\right] . \label{NTILDE}
\end{align}
The above collision integral can be rewritten as 
\begin{align}
\mathcal{I}\left[ \delta n_{\vec{k}}\right] = -\frac{\hbar \vec{k}}{\tau}\cdot \nabla_{\vec{k}} \delta n_{k}
 -\vec{F}^s_{\vec{k}} \cdot \nabla_{\vec{k}} \delta n_{k},
\end{align}
with 
\begin{align} \label{sefB}
  \vec{\mathcal{B}}_s&= -\frac{\hbar k \omega _{s}}{ev_F}\hat{\vec{z}}.
\end{align}
Thus, as anticipated in Sec. \ref{sec:BE} (cf. Eqs.~\eqref{QBE00} and \eqref{eq:fls}),  an effective Lorentz-like force driving the SHE emerges from  skew scattering  with adatom impurities. This Lorentz-like force term needs to be factored out of the collision integral, and the remaining terms are grouped in the dissipative part of the the collision integral, $\mathcal{I}_r\left[n_{\vec{k}}\right]$, which we introduced in Eq.~\eqref{QBE00}, See Sec. \ref{sec:BE} .

\section{Diffusion equations}
In order to derive the diffusion equations that we have employed in the main text, let us first consider  the simpler case
where there is no applied magnetic field and therefore the Larmor frequency vanishes, i.e. $\omega_L=0$ in Eq.~\eqref{QBE0}.

  First of all, let us the define currents and generalized polarization densities as follows:
\begin{align}
 J^\nu_{i}&=\sum_{\vec{k}}ev_i^{\vec{k}}\mathrm{Tr}%
\left[ \gamma ^{\nu}\delta n_{\vec{k}}\right] ,\\
N^{\nu}&=\sum_{\vec{k}}e\mathrm{Tr}\left[ \gamma ^{\nu}\delta n_{\vec{k}}\right].
\end{align}
At zero temperature, $J^\nu_{i} $ and $N^{\nu} $ reduce to:
\begin{align}
 J^{\nu}_{i}\left( \vec{r}\right) &=e\nu_F \mu_F v^{\nu}_{i}\left( \vec{r}\right) ,
\\
 N^{
\nu}\left( \vec{r}\right)& =2e\nu_F \mu ^{\nu}\left( \vec{r}\right) ,
\end{align}
where $\nu_F =k_{F}^{2}/\left( \pi
\mu_F \right) $ is the total density of states at the Fermi energy $\mu_F = \hbar v_F k_F$ at zero temperature, where $k_F$ is the Fermi momentum.

\subsection{Continuity and constitutive equations}

The constitutive and continuity equations in steady state, can be obtained by tracing the linearized QBE \eqref{QBE0}, i.e. by taking $\sum_{\vec{k}}e\vec{v}^{\vec{k}}\text{Tr}\left[ \gamma ^{\mu}\mathrm{QBE} \right] $ for the constitutive equations and $\sum_{\vec{k}}e\text{Tr}\left[ \gamma ^{\mu}\mathrm{QBE} \right]$) for the continuity relations, respectively. The latter procedures yield the following expressions:
\begin{equation} \label{eq:eom_charge}
\mathcal{D} \partial _{i}N^{\mu }- \sigma_D E_{i}^{\mu }=\left[
- \delta ^{\mu}_{\nu}\delta _{ij} +(R_{H})^{\mu}_{\nu}\epsilon
_{ij}\right] J_{j}^{\nu} , 
\end{equation}
\begin{equation} \label{2.20}
\partial_i J^{\mu}_{i}
=0.
\end{equation}
Here $\mathcal{D} = v^2_F \tau/2$ is diffusion constant and $\sigma _{D}=ne^{2}\tau /m$ ($n$ is carrier density and $m$ is mass) is Drude conductivity. In the above expression, repeated indices are summed and $\epsilon _{ij}$ is the antisymmetric 2D Levi-Civita tensor ($i,j = x, y$). The Greek superscripts of the currents  $\vec{J}^{\mu}$ and the densities $N^{\mu}$  take values over the set $\{c,sv,v,s\}$, which stand for for charge, spin-valley, valley, and spin currents (densities), respectively. The coupling between spin and valley currents naturally leads to the existence of spin  and valley polarized currents that are \emph{longitudinal}, i.e. have the same direction as charge current $\vec{J}^{c}$(external electric field $\vec{E}^{c}=\vec{E}$). On the other hand, the spin and valley currents are \emph{transverse}, i.e. perpendicular to $\vec{J}^{c}$ ($\vec{E}$).

The left hand side of Eq.~\eqref{eq:eom_charge} contains the driving forces for the currents, which are the results of spatial nonuniformity of the densities  $\propto \vec{\nabla} N^{\mu}$ and  the application of the generalized electric fields $\vec{E}^{\mu}$ (in order to describe real devices, we shall set $\vec{E}^{\mu} =0$ for all $\mu \neq c$). The right hand side  of Eq.~\eqref{eq:eom_charge} describes the effective Lorentz forces as well as current relaxation. The relaxation rates for all currents are the same and equal to the Drude relaxation time $\tau$ (which is related to the mean-free path by $\ell = v_F \tau$ where $v_F$ is the Fermi velocity). The Hall resistivity matrix $R_{H}$ describes SHE and VHE, and couples \textit{longitudinal} charge and spin-valley currents to \textit{transverse} spin and valley currents:
\begin{align}
R_{H}=
\begin{bmatrix}
0 & 0 & \omega _{v}\tau & \omega _{s}\tau \\
0 & 0 & \omega _{s}\tau & \omega _{v}\tau \\
\omega _{v}\tau & \omega _{s}\tau & 0 & 0 \\
\omega _{s}\tau & \omega _{v}\tau & 0 & 0 \\
\end{bmatrix}.
\end{align}
The magnitude of the SHE and VHE has been parameterized in the above equations by the effective ``cyclotron'' frequencies 
\begin{align}
\omega _{s}&=v_F e \mathcal{B}_s/\hbar k_F\\
\omega_{v}&=v_Fe\mathcal{B}_v/\hbar k_F. 
\end{align}
The latter arise from  effective Lorentz forces that deflect the electrons (according to their spin and valley orientations, respectively).  

In order to describe spin-valley transport with the above equations, we need to invert the resistivity matrix $R_{H}$ and solve Eq.~\eqref{eq:eom_charge} for the currents $J_{i}^{\mu}$, which yields the following set of  equations:
\begin{equation}
J_{i}^{\mu }=-\left(D_{ij}\right)^{\mu}_{\nu }\partial _{j}N^{\nu}+\left(\sigma _{ij}\right)^{\mu}_{\nu
}E_{j}^{\nu }.  \label{eq:J_new}
\end{equation}
Note that the diffusion matrix is a rank-$2$ tensor in the Latin indices $i,j$, and therefore it can be split into a symmetric ($\propto\delta _{ij}$) and antisymmetric ($\propto \epsilon_{ij}$) part according to $ D_{ij}=D_{0}\delta _{ij}+D_{H}\epsilon _{ij}$, where
\begin{equation}  \label{eq:D0}
D_{0}=\mathcal{D}_{r}%
\begin{bmatrix}
1 & \eta  & 0 & 0 \\
\eta  & 1 & 0 & 0 \\
0 & 0 & 1 & \eta  \\
0 & 0 & \eta & 1%
\end{bmatrix},
\end{equation}
\begin{equation} \label{eq:D1}
D_{H}=\mathcal{D}_{r}
\begin{bmatrix}
0 & 0 & \theta_{v} & \theta_{s} \\
0 & 0 & \theta_{s} & \theta_{v} \\
\theta_{v} & \theta_{s} & 0 & 0 \\
\theta_{s} & \theta_{v} & 0 & 0%
\end{bmatrix},
\end{equation}
\begin{equation}  \label{eq:D3}
\mathcal{D}_{r}=\mathcal{D}\frac{1+(\omega_{v}\tau )^{2}+(\omega_{s}\tau )^{2}}{[1+(\omega
_{v}\tau )^{2}+(\omega _{s}\tau )^{2}]^{2}-4\omega_{s}\omega_{v}\tau ^{2}}.
\end{equation}
Similarly, a decomposition of conductivity matrix as 
$\sigma_{ij} = \sigma_0 \delta_{ij} + \sigma_H \epsilon_{ij}$ 
can be obtained by replacing in the above expressions the diffusion constant $\mathcal{D}$ with the Drude conductivity $\sigma_{D}$.  (See exact expressions for $\eta,\theta_v, \theta_s$ in manuscript. )

\subsection{Diffusion of spin and valley polarization}
Next, we derive the drift-diffusion equations for the spin and valley polarizations. To this end,  we supplement the constitutive relations in Eq.~\eqref{eq:eom_charge} with the steady state phenomenological continuity equations,
\begin{equation} 
\partial_i J^{\mu}_{i}
=%
-\frac{\delta ^{\mu}_{\nu}}{ \tau ^{\nu}}
N^{\nu },
\end{equation}
where we take $\tau^c\to +\infty$ since the charge current is strictly conserved. In the above expressions, $\tau^{\mu}$ are phenomenological relaxation times which need to be \emph{ad hoc} in the present derivation, but whose existence can be rigorously derived in a more complete treatment~\cite{huang2016direct, zhang2016valley}.  
Hence, we arrive at the following set of diffusion equations:
\begin{equation}
(D_{0})^{\mu }_{\nu }\partial _{i}^{2}N^{\nu }-\frac{\delta ^{\mu  }_{ \nu }}{%
\tau ^{\nu }}N^{\nu }=S^\nu
\label{eq:dde},
\end{equation}
where the source term is given by
\begin{equation}
S^\nu=\epsilon _{ij}\left[ -\partial _{i}(D_{H})^{\mu }_{\nu }\partial
_{j}N^{\nu }+\partial _{i}(\sigma _{H})^{\mu }_{\nu }E_{j}^{\nu }\right].
\end{equation}
In deriving the
above diffusion equations, we used  $\epsilon _{ij}\partial _{i}\partial _{j}N^{\mu }=0$ and that the generalized electric field is curl and divergence-free, i.e. $\epsilon _{ij}\partial
_{i}E_{j}^{\mu }=0$ and $\partial _{i}E_{i}^{\mu }=0$, that is, we have neglected any relativistic corrections to the electrodynamics.

Note that the source term on the right hand side of~\eqref{eq:dde} takes a non-zero values only at the boundary of the device. In other words,  it
describes the driving force for the electron diffusion arising from the abrupt change of the 
Hall angle at the device boundaries~\cite%
{beconcini2016nonlocal}.  However, in the bulk the above set of differential equations (%
\ref{eq:dde}), becomes a homogeneous one:
\begin{equation} \label{sdq}
\partial_i ^{2}N^{\mu }-\mathcal{M}^{\mu}_{\nu }N^{\nu }=0,  
\end{equation}
where
\begin{equation}  \label{2.66}
\mathcal{M}^{\mu}_{\nu}=\frac{1}{1-\eta ^{2}}%
\begin{bmatrix}
\ell _{v}^{-2} & -\eta \ell _{s}^{-2} \\
-\eta\ell _{v}^{-2} & \ell _{s}^{-2}
\end{bmatrix}. 
\end{equation}
Here $\mu,\nu  \in \{v,s\}$ denote  the transverse valley  (spin) response, with diffusion lengths $\ell _{v}=\sqrt{\mathcal{D}
_{r}\tau ^{v}}$
($\ell _{s}=\sqrt{\mathcal{D}_{r}\tau ^{s}}$).
The choice where $\mu,\nu  = \{c,sv\}$ 
corresponds to the longitudinal charge (spin-valley) response, which decouples from transverse modes and will be omitted in what follows.  The parameter $\eta$, which arises from the interplay of SHE and VHE, mixes the valley and spin responses. 

 As described in the main text, in order to solve Eq.~\eqref{sdq}, we first need to diagonalize the matrix $\mathcal{M}$ and therefore obtain the eigenvalues and eigenvectors. Thus, in what follows we shall assume this has been carried out, so that  $\mathcal{L}
_{a}^{-2}|\vec{\hat{e}}^{\mu}_{a}\rangle =\mathcal{M}^{\mu }_{ \nu }|\vec{\hat{e}}%
^{\nu}_{ a}\rangle $, where  $\mathcal{L}_a$ is the eigenvalue, which corresponds to the diffusion length for the eigenmode $|\vec{\hat{e}}^{\mu}_{a}\rangle$.

Next, following Beconcini~\emph{et al.}~\cite{beconcini2016nonlocal}, we solve
 the diffusion equation for a Hall bar device geometry, assuming the latter to be an infinitely long  metallic channel of width $w$ contacted by noninvasive current and voltage probes (see Fig. 1(a) in the manuscript). We  shall assume the
complete screening of the electric field in the bulk of device, which amounts to
take charge density into zero, i.e., $N^{c}\left( \vec{r}\right)
=0.$ Hence, the electrostatic potential, $\Phi \left( \vec{r}\right)$ obeys the Laplace equation: 
\begin{equation} \label{2.6}
\nabla ^{2}\Phi \left( \vec{r}\right) =0. 
\end{equation}
The Laplace equation  \eqref{2.6} and the above system of partial differential equations \eqref{sdq}, need to be
supplemented by the following boundary conditions (BCs): 
\begin{align}
J^{c}_{y}(x;y=\pm w/2)&=I\delta (x) \\
J^{\nu}_{y}\left( x;y=\pm w/2\right) &=0,
\end{align}
for $\nu=v,s$. $I$ is charge current injected on right hand side of Hall bar device. Finally, in order to solve the problem posed by Eq.~\eqref{sdq} and
Eq.~\eqref{2.6}, we use  Fourier transformation along the infinitely long channel direction, $x$.
Thus,  using~\eqref{eq:J_new}, the BCs
approximately become: 
\begin{equation}
I \simeq \left. \left[-
\mathcal{D}_{r} \left( ik\right) N
^{\nu}\left( k,y\right)  \theta _{\nu}   -\sigma ^{c}\partial _{y}\Phi \left( k,y\right)\right]\right\vert _{y=\pm \frac{w}{2}},  \label{2.11A}
\end{equation}
where the sum over the repeated index $\nu$ in the expression above runs over the set $\{s,v\}$ only. $\sigma^c$ is charge conductivity. In addition,
\begin{equation}
0\simeq \left. \left[-\sigma^{c} \left( ik\right) \Phi
\left( k,y\right) \theta _{\nu } -\mathcal{D}_{r}\partial _{y}N^{\nu }\left(
k,y\right)\right] \right\vert _{y=\pm \tfrac{w}{2}}. \label{2.11B}
\end{equation}
By ``approximately'', we mean that we omit the boundary contributions of the longitudinal modes $N^{c}$ and $N^{sv}$ in Eq. (\ref{eq:J_new}) by setting $N^c, N^{sv} = 0$ and $\eta = 0$. Including them, merely leads to a small correction to the diffusion length of the spin and valley eigenmodes.

In order to solve the above system of 2nd order differential equations, i.e. Eq.~\eqref{sdq}, we first turn it into a 1st order set of equations by defining $N_{\nu }^{\prime }\left( k,y\right)
=\partial _{y}N^{\nu }\left( k,y\right)$,  rendering~\eqref{sdq} to the form:
\begin{equation}
\begin{bmatrix}
\partial _{y}N^{\nu }\left( k,y\right) \\ 
\partial _{y}N_{\mu }^{\prime }\left( k,y\right)%
\end{bmatrix}%
=%
\begin{bmatrix}
0 & \delta^{\nu}_{\mu} \\ 
k^{2}\delta^{\mu}_{\nu}+\mathcal{M}^{\mu}_{\nu} & 0%
\end{bmatrix}%
\begin{bmatrix}
N^{\nu }\left( k,y\right) \\ 
N_{\mu }^{\prime }\left( k,y\right)%
\end{bmatrix}%
.  \label{2.9A}
\end{equation}
Let $\mathcal{L}
_{a}^{-2} $ and $|\vec{\hat{e}}^{\mu}_{a}\rangle$ are the eigenvalues and eigenvectors of diffusion matrix, respectively. Hence,
\begin{equation}
\begin{bmatrix}
\vec{0} & \mathbb{1} \\ 
k^{2}\mathbb{1}+\mathcal{M} & \vec{0}%
\end{bmatrix}%
\begin{bmatrix}
\left\vert \vec{\hat{e}}_{a }\right\rangle \\ 
\pm \kappa _{a }\left\vert \vec{\hat{e}}_{a }\right\rangle%
\end{bmatrix}%
=\pm \kappa _{a }%
\begin{bmatrix}
\left\vert \vec{\hat{e}}_{a }\right\rangle \\ 
\pm \kappa _{a }\left\vert \vec{\hat{e}}_{a }\right\rangle%
\end{bmatrix}%
,
\end{equation}%
with $\kappa _{a}=\sqrt{k^{2}+\mathcal{L}
_{a}^{-2}}$. Therefore,  $\pm
\kappa _{a}$ is the eigenvalue of the matrix of \eqref{2.9A} with eigenvector 
\begin{equation}
\left\vert \pm \kappa _{a }\right\rangle =\frac{1}{(1+|\kappa _{a
}|^{2})^{1/2}}
\begin{bmatrix}
\left\vert e_{a }\right\rangle \\ 
\pm \kappa _{a }\left\vert e_{a }\right\rangle%
\end{bmatrix}.
\end{equation}

Considering the symmetry of BCs in~\eqref{2.11A} and \eqref{2.11B}, the solution of the above system of differential can be solved by the following ansatz:
\begin{equation} \label{2.12A}
N^{\nu }=\sum_{a=1,2}A_{a }\vec{\hat{e}}^{\nu }_{a }\left(
e^{+\kappa _{a }y}+e^{-\kappa _{a }y}\right) ,  
\end{equation}%
\begin{equation} \label{2.12B}
\Phi =A_{o}\left( e^{+ky}-e^{-ky}\right) .  
\end{equation}
Substitution of these ansatz into the BCs, Eq.~\eqref{2.11A} and \eqref{2.11B} yields:
\begin{widetext}
\begin{equation}
I=-\sigma ^{c}\left[ A_{o}k\left( e^{+kw/2}+e^{-kw/2}\right) %
\right] -\mathcal{D}_{r}\sum_{\nu ,a }\theta _{\nu }\left( ik\right) %
\left[ A_{a }\vec{\hat{e}}^{\nu}_{ a }\left( e^{+\kappa _{a }w/2}+e^{-\kappa _{a
}w/2}\right) \right] ,  \label{HS}
\end{equation}
\begin{equation}
0\simeq -\sigma ^{c}\theta _{\nu }\left( ik\right) A_{o}\left(
e^{+kw/2}-e^{-kw/2}\right) -\mathcal{D}_{r}\sum_{a }A_{a }\vec{\hat{e}}%
^{\nu}_{a}\kappa _{a }\left( e^{+\kappa _{a }w/2}-e^{-\kappa _{a
}w/2}\right).  \label{ES}
\end{equation}
From Eq.~\eqref{ES}, it is found that
\begin{equation}
\frac{A_{a }}{A_{o}}\simeq \left( -i\right) \frac{\sigma ^{c}}{%
\mathcal{D}_{r}}\sum_{\mu }\frac{k\sinh \left( kw/2\right) }{\kappa _{a
}\sinh \left( \kappa _{a }w/2\right) }(\vec{\hat{e}}^{-1})_{a}^{\mu }\theta _{\mu
},
\end{equation}
Hence, upon substitution of this result into Eq.~\eqref{HS}, we obtain:
\begin{equation}
\frac{I}{A_{o}}=-2k\cosh \left( k\frac{W}{2}\right) \left[ 1+\sum_{a
}\Theta _{a }^{2}\mathcal{F}_{a}\left( k\right) \right] \sigma
^{c},
\end{equation}
where
\begin{equation}
\Theta _{a }^{2}=[\theta _{\mu }\vec{%
\hat{e}}^{\mu}_{ a}][(\vec{\hat{e}}^{-1})^{a}_{\nu }\theta _{\nu }],
\end{equation}%
\begin{equation}
\mathcal{F}_{a}\left( k\right) =\frac{k\tanh \left( kw/2\right) }{\kappa
_{a}\tanh \left( \kappa _{a }w/2\right) }.
\end{equation}
Hence,
\begin{equation} \label{Generalized density}
N^{\nu }\left( \vec{r}\right) =\frac{iI}{\mathcal{D}_{r}}\sum_{a ,\mu }\vec{\hat{e}}^{\nu}_{ a }(\vec{\hat{e}}^{-1})^{a}_{\mu }\theta _{\mu } \int dk%
\frac{e^{ikx}}{2\pi }\frac{\tanh \left( kw/2\right) }{\kappa _{a }\sinh
\left( \kappa _{a }w/2\right) }\frac{\cosh \left( \kappa _{a }y\right) }{%
\left[ 1+\sum_{b }\Theta _{b }^{2}\mathcal{F}_{b }\left(
k\right) \right] }, 
\end{equation}
for the generalized polarization densities and
\begin{equation} \label{Scalar potential}
\Phi \left( \vec{r}\right) =-\frac{I}{\sigma ^{\mathrm{c}}}\int dk\frac{e^{ikx}}{2\pi k}%
\frac{\sinh \left( ky\right) }{\cosh \left( kW/2\right) \left[ 1+\sum_{b }\Theta _{b }^{2}\mathcal{F}_{b }\left(
k\right) \right] }.
\end{equation}
\end{widetext}
for the electrostatic potential.

\section{Nonlocal Resistance}

In this section, we compute the nonlocal resistance (NLR) in the absence of magnetic field, which is defined as 
\begin{equation}
R_{nl}\left( x\right)
= \frac{1}{I} [\Phi \left( x,-w/2\right) -\Phi \left( x,+w/2\right)]. \label{eq.rnl}
\end{equation}
Substituting the electrostatic potential~\eqref{Scalar potential} into~\eqref{eq.rnl}, we obtain the  following integral form for the NLR:
\begin{equation}
\frac{R_{nl}(x)}{R_{\mathrm{xx}}}=\frac{1}{\pi }\int_{-\infty
}^{\infty }dk\frac{e^{ik x}}{k}\frac{\tanh \left( k w/2\right) }{1+\sum_{b
}\Theta _{b }^{2}\mathcal{F}_{b}\left( k\right) },
\end{equation}
with $R_{\mathrm{xx}}=1/\sigma^c$. The above result for the NLR can be expanded as follows
\begin{align}
\frac{R_{nl}(x)}{R_{\mathrm{xx}}}=\sum_{n=0}^{\infty }\mathcal{R}%
^{n}\left( x\right), \\
\mathcal{R}^{n}\left( x\right) =\sum_{\vec{a}^n}%
\mathcal{R}_{\vec{a}^{n}}\left( x\right) ,
\end{align}
with $\vec{a}^n=(a_1,a_2,\cdots,a_n)$. The expression for $\mathcal{R}_{\vec{a}^n}\left( x\right)$ is given by:
\begin{equation}
\mathcal{R}_{\vec{a}^n}(x)=\frac{1}{\pi }\left( -1\right) ^{n}\int_{-\infty
}^{+\infty }\frac{dk\, e^{ik x}}{k\coth \left( k w/2\right) }\prod\limits_{\imath
=1}^{n}\Theta _{a_{\imath }}^{2}\mathcal{F}_{a _{\imath }}\left(
k\right) .
\end{equation}
Next, we obtain asymptotic expressions for the various terms in the above expansion. For $n=0$, $\mathcal{R}^{0}\left(x\right) $ reduces to the Ohmic NLR:
\begin{equation}
\mathcal{R}^{0}(x)=\frac{2}{\pi }\ln \left\vert \coth \left( \frac{\pi x}{2w}%
\right) \right\vert.
\end{equation}
Explicitly, it is van der Paw resistance, which behaves as $\mathcal{R}_{\mathrm{%
vdP}}\simeq \frac{4}{\pi }e^{-\left\vert x\right\vert /\mathcal{L}_{0}}$ for 
$\left\vert x\right\vert \gg w$ where $\mathcal{L}_{0}=w/\pi $. At large 
$\left\vert x\right\vert $, and for $w\ll \ell _{\nu }$, the $n=1$ term is $%
\mathcal{R}^{1}=\sum_{a }\mathcal{R}_{a }^{1},$ 
where
\begin{equation}
\mathcal{R}_{a }^{1}\left( x\right) \simeq \Theta _{a }^{2}\frac{w}{%
2\mathcal{L}_{a }}e^{-\left\vert x\right\vert /\mathcal{L}_{a }}.
\end{equation}
In earlier work~\cite{zhang2016valley}, we showed that a modest nonuniform strain can result in rather large  valley Hall angles $\theta_v \sim 1$. Thus, in order to accurately describe the NLR we need to
consider high order terms in the expansion, i.e. those with $n>1$. But we here just pick out terms $\mathcal{R}_{\vec{a}^{n}}$ with same eigenmode $a_{\imath }=a $
i.e., $R_{nl}/R_{\mathrm{xx}}\simeq \mathcal{R}^{0}+\sum_{a}%
\mathcal{R}^{a }$, being
\begin{align}
\mathcal{R}^{a}(x) &=\int_{-\infty }^{+\infty }\frac{dk}{\pi k}\frac{e^{ikx}%
}{\coth \left( k w/2\right) }\sum_{n=1}^{\infty }\left( -1\right) ^{n}\left[
\Theta _{a }^{2}\mathcal{F}_{a }\left( k\right) \right] ^{n}\notag  \\
&=\frac{\Theta _{a }^{2}}{1+\Theta _{a }^{2}}\frac{W}{2\mathcal{L}^r_{a
}}e^{-\left\vert x\right\vert /\mathcal{L}^r_{a }},  
\end{align}
where $\mathcal{L}^r_{a }=\sqrt{1+\Theta _{a }^{2}}\mathcal{L} _{a }$ is
renormalized decay lengths of each eigenmode~\cite{zhang2016valley}.
Finally, we obtain total NLR $R_{nl}/R_{\mathrm{xx}}=\mathcal{R}%
^{0}+\delta \mathcal{R}_{nl}$ 
\begin{equation}
\frac{R_{nl}(x)}{R_{\mathrm{xx}}}\simeq \overbrace{\frac{4}{\pi }%
e^{-\left\vert x\right\vert /\mathcal{L}_{0}}}^{\mathcal{R}^{0}}+\overbrace{%
\sum_{a }\underbrace{\frac{\Theta _{a }^{2}}{1+\Theta _{a }^{2}}\frac{w%
}{2\mathcal{L}^r_{a }}e^{-\left\vert X\right\vert /\mathcal{L}^r_{a}}}_{\mathcal{R}^{a }}}^{\delta \mathcal{R}_{nl}},  \label{NLR}
\end{equation}
The first term is the Ohmic contribution, $\mathcal{R}^{0}$, and the second term
contains the sum  of the exponentially decaying contributions for each eigenmode, $%
\mathcal{R}^{a }$. Near the current injection point ($\left\vert
x\right\vert \lesssim \mathcal{L}_{0}$), $R_{nl}$ is dominated by the
ohmic contribution, $\mathcal{R}^{0}$, which will become negligible at sufficiently large
distances (i.e. for $\left\vert x\right\vert \gg \mathcal{L}_{0}$). 

Here we focus on the behavior of $R_{nl}$, when contribution of the eigenmodes of the diffusion equation dominate over the Ohmic contribution, i.e. when $\delta \mathcal{R}_{nl}\gg \mathcal{R}^{0}$.

\section{Suppression of the Hanle effect}
In this section, we provide the details of the derivation and solution of the diffusion equations in the presence of an in-plane magnetic field. Note that the Larmor frequency $\omega_L\ll \mu_F$, where $\mu_F$ is the Fermi level. The in-plane magnetic field, which we shall take parallel to the direction of the electric field applied to the device, induces precession of the spin-degree of freedom, whilst the valley is not affected. This mixes the out-of-plane spin component along $z$ with the spin in-plane components along the $x$ and $y$ axes. Thus, our ansatz for the density-matrix distribution function in the QBE must be now expanded in terms of $\gamma_{\nu}$ matrices taken from the larger set  
$\{\hat{s}_{o}\hat{\tau}_{o},\hat{s}_{o}\hat{\tau}%
_{z},\hat{s}_{x}\hat{\tau}_{o},\hat{s}_{y}\hat{\tau}_{o},\hat{s}_{z}\hat{\tau%
}_{o},\hat{s}_{x}\hat{\tau}_{z},\hat{s}_{y}\hat{\tau}_{z},\hat{s}_{z}\hat{%
\tau}_{z}\}$.

 In order to simplify the calculations described below, the deviation of the distribution function
from equilibrium, i.e. $\delta n_{\vec{k}} = n_{\vec{k}}-n_{\vec{k}
}^{0}$,  will be slit into  two parts, $\delta n_{\vec{k}}=\delta n_{\vec{k}}^{+}+\delta n_{\vec{k}}^{-}$, with
\begin{equation}
\delta n_{\vec{k}}^{+}\simeq \sum_{\mathrm{i}}\gamma_{\mathrm{i}}\left[ \mu
^{\mathrm{i}}\left( \vec{r}\right) +\vec{v}^{\mathrm{i}}\left( \vec{r}%
\right) \cdot \hbar \vec{k}\right] \left[ -\partial _{\epsilon }n^{0}\left(
\epsilon \right) \right] _{\epsilon =\mu_F },  \label{3.25}
\end{equation}
\begin{equation}
\delta n_{\vec{k}}^{-}\simeq \sum_{\mathrm{j}}\gamma_{\mathrm{j}}\left[ \mu
^{\mathrm{j}}\left( \vec{r}\right) +\vec{v}^{\mathrm{j}}\left( \vec{r}%
\right) \cdot \hbar \vec{k}\right] \left[ -\partial _{\epsilon }n^{0}\left(
\epsilon \right) \right] _{\epsilon =\mu_F },  \label{3.26}
\end{equation}
where $\mathrm{i} \in \left\{c,sv,v,s\right\} $ and $\mathrm{j}\in \left\{ xo,yo,xz,yz
\right\}$. 

The  form of the  collision integral~\eqref{E.222} is determined by  the ansatz for
density matrix $\delta n_{\vec{k}}$, which in turn follows from the forms of the $T$-matrix ($\propto\{\hat{s}_{o},\hat{s} _{z}\}$),  the pseudo-magnetic field arising from nonuniform strain ($\propto \{\hat{\tau}_{o},\hat{\tau}_{z}\}$), and the (Zeeman) 
magnetic field ($\{\hat{s}_{o},\hat{s}_{y}\}$). To compute the collision integral, it  is convenient to also split the $T$-matrix into two parts, i.e.,  $T_{\vec{k}\vec{p}}=T_{\vec{k}\vec{p}}^{o}+T_{%
\vec{k}\vec{p}}^{z}$, with 
\begin{align}  \label{TM}
T_{\vec{k}\vec{p}}^{o}&=t_{c}\mathbb{1}\cos \left( \frac{%
\theta }{2}\right) ,\\
T_{\vec{k}\vec{p}}^{z}&=it_{s}\hat{s}_{z}\sin
\left( \frac{\theta }{2}\right) , 
\end{align}
which obey:
\begin{align}
\left[ \delta n_{\vec{k}}^{-},T_{\vec{k}\vec{p}}^{z}\right] _{+}&=0,\\
\left[
\delta n_{\vec{k}}^{-},T_{\vec{k}\vec{p}}^{o}\right] _{-}&=0,\\
\left[
\delta n_{\vec{k}}^{+},T_{\vec{k}\vec{p}}\right] _{-}&=0,
\end{align}
where $\left[ A,B\right] _{\pm }= AB\pm BA$. Next, using the above
ansatz, the collision integral (\ref{E.222}) reduces to:
\begin{align} \label{3.28888}
\mathcal{I}\left[ \delta n_{\vec{k}}\right] &=\frac{\pi }{\hbar }n_{\mathrm{
i}}\sum_{\vec{p}}\delta \left[ \epsilon \left( p\right) -\epsilon \left(
q\right) \right] 2\hat{T}_{\vec{k}\vec{p}}\hat{T}_{\vec{k}\vec{p}}^{\ast
}(\delta n_{\vec{p}}^{+}-\delta n_{\vec{k}}^{+})   \\
&+\frac{\pi }{\hbar }n_{\mathrm{i}}\sum_{\vec{p}}\delta \left[ \epsilon
\left( p\right) -\epsilon \left( q\right) \right] 2\hat{T}_{\vec{k}\vec{p}}%
\hat{T}_{\vec{k}\vec{p}}^{o\ast }(\delta n_{\vec{p}}^{-}-\delta n_{\vec{k}%
}^{-})  \notag \\
&-\frac{\pi }{\hbar }n_{\mathrm{i}}\sum_{\vec{p}}\delta \left[ \epsilon
\left( p\right) -\epsilon \left( q\right) \right] 2\hat{T}_{\vec{k}\vec{p}}%
\hat{T}_{\vec{k}\vec{p}}^{z\ast }(\delta n_{\vec{p}}^{-}+\delta n_{\vec{k}%
}^{-}).  \notag
\end{align}
Hence,
\begin{align} \label{TT}
2\hat{T}_{\vec{k}\vec{p}}^{+}\hat{T}_{\vec{k}\vec{p}}^{\ast } &=\left[
\left\vert t_{c}\right\vert ^{2}\left( 1+\cos \theta \right)
+\left\vert t_{s}\right\vert ^{2}\left( 1-\cos \theta \right) %
\right] \gamma ^{c}\notag  \\
&+2\text{\textrm{Im}}\left( t_{c}t_{s}^{\ast }\right)
\sin \theta \gamma ^{s} 
\end{align}%
\begin{equation}
2\hat{T}_{\vec{k}\vec{p}}\hat{T}_{\vec{k}\vec{p}}^{o\ast }=\left\vert t_{%
c}\right\vert ^{2}\left( 1+\cos \theta \right) \gamma ^{c%
}+it_{s}t_{c}^{\ast }\sin \theta \gamma ^{s},
\end{equation}%
\begin{equation}
2\hat{T}_{\vec{k}\vec{p}}\hat{T}_{\vec{k}\vec{p}}^{z\ast }=\left\vert t_{%
s}\right\vert ^{2}\left( 1-\cos \theta \right) \gamma ^{c%
}-it_{c}t_{s}^{\ast }\sin \theta \gamma ^{s},
\end{equation}
In addition, we need to compute the differences and sums:
\begin{equation}
\delta n_{\vec{p}}^{-}-\delta n_{\vec{k}}^{-}=\left[ -\partial _{\epsilon
}n^{0}\left( \epsilon \right) \right] _{\epsilon =\mu_F }\sum_{\mathrm{j}%
}\gamma _{\mathrm{j}}\vec{v}_{\mathrm{j}}\left( \vec{r}\right) \cdot \hbar (%
\vec{p}-\vec{k}),
\end{equation}%
\begin{equation}
\delta n_{\vec{p}}^{+}-\delta n_{\vec{k}}^{+}=\left[ -\partial _{\epsilon
}n^{0}\left( \epsilon \right) \right] _{\epsilon =\mu_F }\sum_{\mathrm{i}%
}\gamma _{\mathrm{i}}\vec{v}_{\mathrm{i}}\left( \vec{r}\right) \cdot \hbar (%
\vec{p}-\vec{k}),
\end{equation}%
\begin{align} \label{NN}
\delta n_{\vec{p}}^{-}+\delta n_{\vec{k}}^{-} &=\left[ -\partial _{\epsilon
}n^{0}\left( \epsilon \right) \right] _{\epsilon =\mu_F }\sum_{\mathrm{j}%
}\gamma _{\mathrm{j}}\vec{v}_{\mathrm{j}}\left( \vec{r}\right) \cdot \hbar (%
\vec{p}+\vec{k})  \notag \\
&+\left[ -\partial _{\epsilon }n^{0}\left( \epsilon \right) \right]
_{\epsilon =\mu_F }\sum_{\mathrm{j}}2\gamma _{\mathrm{j}}\mu _{\mathrm{j}%
}\left( \vec{r}\right). 
\end{align}%
Substituting Eqs.~\eqref{TT}-\eqref{NN} into the collision integral~\eqref{3.28888}, the explicit form of the collision integral is split into three contributions: $\mathcal{I}\left[ \delta n_{\vec{k}}\right] =\mathcal{I}^{+}\left[
\delta n_{\vec{k}}\right] +\mathcal{I}^{0}\left[ \delta n_{\vec{k}}\right] +%
\mathcal{I}^{-}\left[ \delta n_{\vec{k}}\right]$, with
\begin{align} \label{3.30}
\mathcal{I}^{+}\left[ \delta n_{\vec{k}}\right] &=\left[ \partial
_{\epsilon }n^{0}\left( \epsilon \right) \right] _{\epsilon =\mu_F }\hbar \vec{%
k}\cdot \left\{ \frac{\gamma ^{c}}{\tau }\sum_{%
\mathrm{i}}\gamma^{\mathrm{i}}\vec{v}_{\mathrm{i}}\left( \vec{r}\right)
\right.  \notag \\
&+\left.\gamma ^{s}\omega _{s} \sum_{\mathrm{i}}\gamma
^{\mathrm{i}}\vec{v}_{\mathrm{i}}\left( \vec{r}\right) \times \hat{\vec{z}}\right\},
\end{align}%
\begin{equation}
\mathcal{I}^{0}\left[ \delta n_{\vec{k}}\right] =\left[ \partial _{\epsilon
}n^{0}\left( \epsilon \right) \right] _{\epsilon =\mu_F }\frac{\gamma ^{%
c}}{\tau _{s,xy}}\sum_{\mathrm{j}}\gamma ^{\mathrm{j}}\mu _{%
\mathrm{j}}\left( \vec{r}\right) ,  \label{3.31}
\end{equation}%
\begin{align}
\mathcal{I}^{-}\left[ \delta n_{\vec{k}}\right] &=\left[ \partial
_{\epsilon }n^{0}\left( \epsilon \right) \right] _{\epsilon =\mu_F }\hbar \vec{%
k}\cdot \left\{ \frac{\gamma ^{c}}{\tilde{\tau}}\sum_{%
\mathrm{j}}\gamma ^{\mathrm{j}}\vec{v}_{\mathrm{j}}\left( \vec{r}\right)
\right.  \notag \\
&+\left. i\gamma ^{s} \omega _{s} \sum_{\mathrm{j}%
}\gamma ^{\mathrm{j}}\vec{v}_{\mathrm{j}}\left( \vec{r}\right) \times \hat{%
\vec{z}}\right\} ,  \label{3.32}
\end{align}%
where $\mathrm{i}=\left(c,sv,v,s%
\right) $, $\mathrm{j}=\left( xo,yo,xz,yz%
\right) $ and we define other two kinds of relaxation times to describe the
collision of electrons:
\begin{align}
\frac{1}{\tau _{s,xy}(k)}&=\frac{k n_{\mathrm{i}}}{4\hbar ^{2}v_{F}}\left[
4\left\vert t_{s}(k)\right\vert ^{2}\right] ,\\
\frac{1}{\tilde{\tau}(k)}&=\frac{k n_{\mathrm{i}}}{4\hbar ^{2}v_{F}}\left[
\left\vert t_{c}(k)\right\vert ^{2}+\left\vert t_{s}(k)\right\vert ^{2}\right] , \label{TILDE}
\end{align}%
Notice that, in the presence of an in-plane magnetic field the term $\mathcal{I}^{0}\left[ \delta n_{\vec{k}}\right] $ in the colission integral introduces an additional relaxation time,  $\tau _{s,xy}(k)$.

In addition to spin  (spin-valley) current, the in-plane magnetic field couples the out-of-plane and in-plane components of the spin current, $\vec{J}^{xo}$ and $\vec{J}^{yo}$ (spin-valley currents, $\vec{J}^{xz}$ and $\vec{J}^{yz}$). Here we take the magnetic field to be parallel to the applied electric field, i.e. $\vec{H}\parallel \vec{\hat{y}}$,  and  thus the following generalized density $N$ and current $\vec{J}$ appear in our diffusion equations: 
\begin{equation}
\vec{J}=%
\begin{bmatrix}
\vec{J}_{\Vert} \\ 
\vec{J}_{\perp}%
\end{bmatrix}%
,\vec{J}_{\Vert}=%
\begin{bmatrix}
\vec{J}^{c} \\ 
\vec{J}^{sv} \\ 
\vec{J}^{xz} \\ 
\vec{J}^{yo}%
\end{bmatrix}%
,\vec{J}_{\perp}=%
\begin{bmatrix}
\vec{J}^{v} \\ 
\vec{J}^{s} \\ 
\vec{J}^{xo} \\ 
\vec{J}^{yz}%
\end{bmatrix}%
,  \label{3.37}
\end{equation}%
\begin{equation}
N=%
\begin{bmatrix}
N_{\Vert} \\ 
N_{\perp}%
\end{bmatrix}%
,N_{\Vert}=%
\begin{bmatrix}
N^{c} \\ 
N^{sv} \\ 
N^{xz} \\ 
N^{yo}%
\end{bmatrix}%
,N_{\perp}=%
\begin{bmatrix}
N^{v} \\ 
N^{s} \\ 
N^{xo} \\ 
N^{yz}%
\end{bmatrix}%
,  \label{3.38}
\end{equation}%
where we have divided the longitudinal and transverse modes. 
Let us first focus on the continuity equations. In the steady state, they read:
\begin{equation} \label{3.39}
\partial_i J^{\mu}_i=-(\tau^{-1}_{sr})^{\mu}_{\nu}N^{\nu}+\omega^{\mu}_{\nu}
N^{\nu},
\end{equation}
which is obtained by tracing the linearized QBE, i.e. taking $\frac{g_{s}g_{v}}{4}\sum_{\vec{k}}e\text{Tr}\left[ \gamma
^{\mu}\left( \text{QBE}\right) \right]$.
In the above expression
\begin{equation}
\tau^{-1}_{sr}=
\begin{bmatrix}
\tau^{-1}_0 & 0 \\ 
0 & \tau^{-1}_0  
\end{bmatrix},
\tau^{-1}_0=
\begin{bmatrix}
0 & 0 & 0 & 0 \\ 
0 & 0 & 0 & 0 \\ 
0 & 0 & \tau _{s,xy}^{-1} & 0 \\ 
0 & 0 & 0 & \tau _{s,xy}^{-1}%
\end{bmatrix},
\end{equation}%
\begin{equation}
\omega=
\begin{bmatrix}
\omega_0 & 0 \\ 
0 & \omega_0  
\end{bmatrix},
\omega_0=
\begin{bmatrix}
0 & 0 & 0 & 0 \\ 
0 & 0 & +\omega _{L} & 0 \\ 
0 & -\omega _{L} & 0 & 0 \\ 
0 & 0 & 0 & 0%
\end{bmatrix}.
\end{equation}
The matrix $\tau^{-1}_{sr}$ describes the spin relaxation for spin polarized in the $x$-$y$ plane. In our microscopic model, $s_z$ is a good quantum number and there is no relaxation. The second term describes the spin precession induced by an in-plane magnetic field in $y$-axis direction.  
We parameterize the strength of the in-plane magnetic field $H$ by the Larmor frequency 
$\omega _{L}= \textsl{g}\mu_B H
/\hbar$.

The constitutive relations for the generalized currents, $J^{\mu}_i$ is given by following equations:
\begin{equation} \label{CE11111}
\mathcal{D}  \partial _{i}N^{\mu }- \sigma_D E_{i}^{\mu }=\left[
- \delta ^{\mu}_{\nu}\delta _{ij} +\tau\omega ^{\mu}_{\nu}\delta _{ij}+(R_{H})^{\mu}_{\nu}\epsilon
_{ij}\right] J_{j}^{\nu} , 
\end{equation}
where
\begin{equation}
R_{H}=
\begin{bmatrix}
0 & R^0_{H} \\ 
R^0_{H} & 0  
\end{bmatrix},
R^0_{H}=%
\begin{bmatrix}
 \omega _{v}\tau & \omega _{s}\tau & 0 & 0 \\
\omega _{s}\tau & \omega _{v}\tau  & 0 & 0   \\
 0 & 0 & \omega _{v}\tau & \omega _{s}\tau  \\
 0 & 0 & \omega _{s}\tau & \omega _{v}\tau 
\end{bmatrix}.
\end{equation}
For the sake of simplicity, we have assumed that the relaxation rates for all currents are the same and equal to Drude relaxation time ($\tilde{\tau} \simeq \tau$) (See expressions for $\tilde{\tau}$ in Eq. (\ref{TILDE}) for $i=(xo,yo,xz,yz)$ and $\tau$ in Eq. (\ref{NTILDE}) for $i=(c,sv,v,s)$. Thus, we take $\mathcal{D}=v^2_F\tau/2$ ($\sigma _{D}=ne^{2}\tau /m$) to be the same for all types of currents. These assumptions can be relaxed, and will not alter our conclusions qualitatively. $R_{H}$ is the coupling matrix that couples the different currents with each other due to the local impurities and the strain pseudo-magnetic field.  

Solving the constitutive equations \eqref{CE11111} for the currents $J^{\mu}_i$ we obtain: 
\begin{equation} \label{eq:J_neww}
J_{i}^{\mu }=-(D_{ij})^{\mu}_{\nu }\partial _{j}N^{\nu }+(\sigma _{ij})^{\mu}_{\nu
}E_{j}^{\nu }.  
\end{equation}
As pointed out in the main text, the diffusion matrix is a rank-$2$ tensor in the space indices $i,j = x, y$, and therefore it can be split into a symmetric ($\propto\delta _{ij}$) and antisymmetric ($\propto \epsilon_{ij}$) parts according to $
D_{ij}=D_{0}\delta _{ij}+D_{H}\epsilon _{ij}$ where
\begin{equation}
D_{0}=
\begin{bmatrix}
D^0_{0} & 0 \\ 
0 & D^0_{0}  
\end{bmatrix},
D^0_{0}= \mathcal{D}_{r}%
\begin{bmatrix}
\eta _{c} & \eta _{sv} & \eta _{xz} & \eta _{yo} \\ 
\eta _{sv}  & \eta _{c} & \eta _{yo} & \eta _{xz} \\ 
\eta _{xz} & \eta _{yo} & \eta _{c} & \eta _{sv} \\ 
\eta _{yo} & \eta _{xz} & \eta _{sv} & \eta _{c}%
\end{bmatrix}%
,
\end{equation}
\begin{equation}
D_{H}=
\begin{bmatrix}
0 & D^0_{H} \\ 
D^0_{H} & 0  
\end{bmatrix},
D^0_{H}=\mathcal{D}_{r}%
\begin{bmatrix}
\theta _{v} & \theta _{s} & \theta _{xo} & \theta _{yz} \\ 
\theta _{s} & \theta _{v} & \theta _{yz} & \theta _{xo} \\ 
\theta _{xo} & \theta _{yz} & \theta _{v} & \theta _{s} \\ 
\theta _{yz} & \theta _{xo} & \theta _{s} & \theta _{v}%
\end{bmatrix}.%
\end{equation}
$\mathcal{D}_{r}, \eta_{\mu}, \theta_{\mu} $ are rather complicated functions of $\omega_v\tau, \omega_s\tau$ and $\omega_L\tau$, and are not given here. 
Similarly, the conductivity matrix can be obtained by replacing the diffusion constant $\mathcal{D}$ with the Drude conductivity $\sigma _{D}$.

\begin{figure}[t]
\begin{center}
\includegraphics[width=0.40\textwidth]{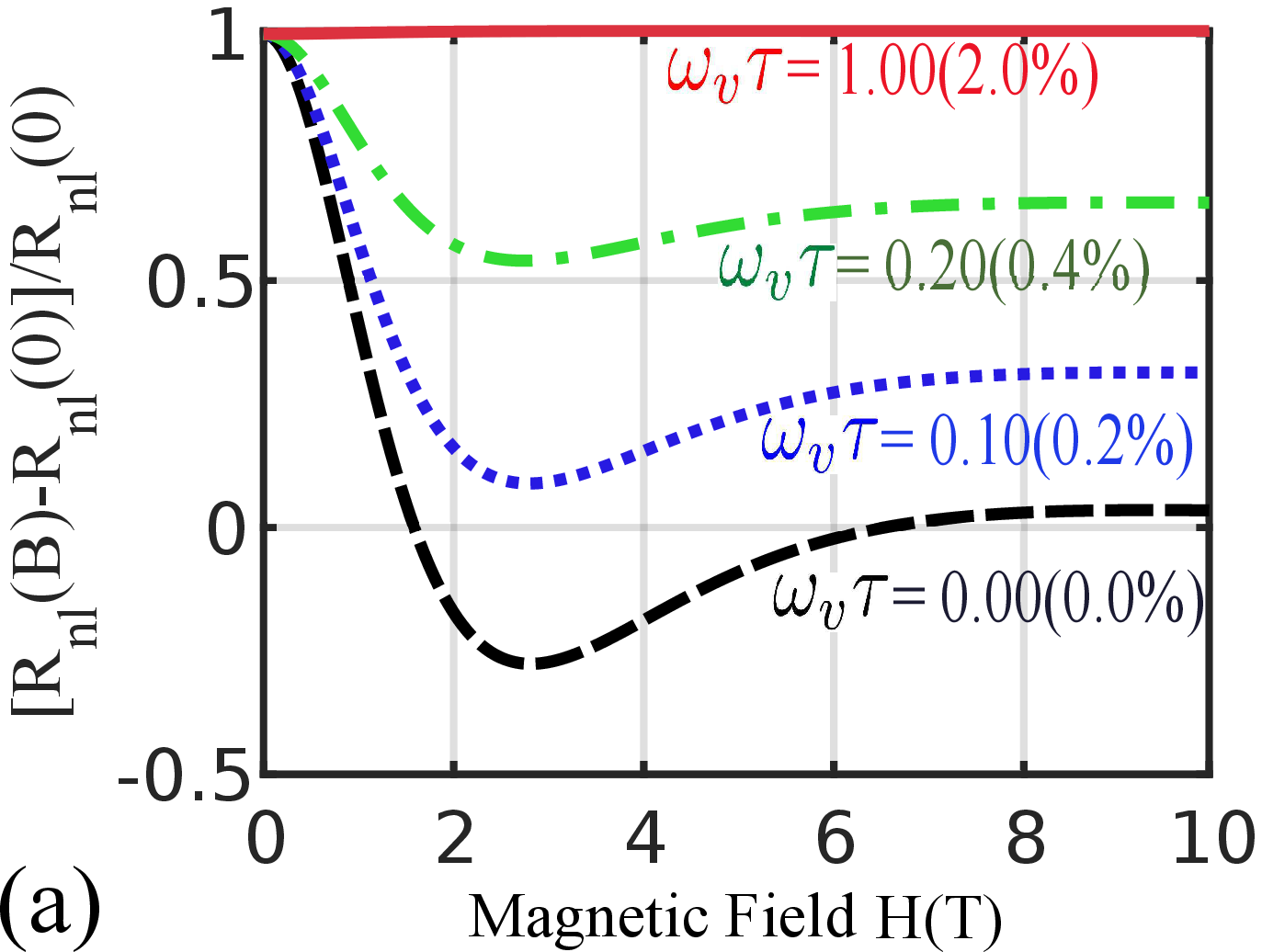}\\
\includegraphics[width=0.40\textwidth]{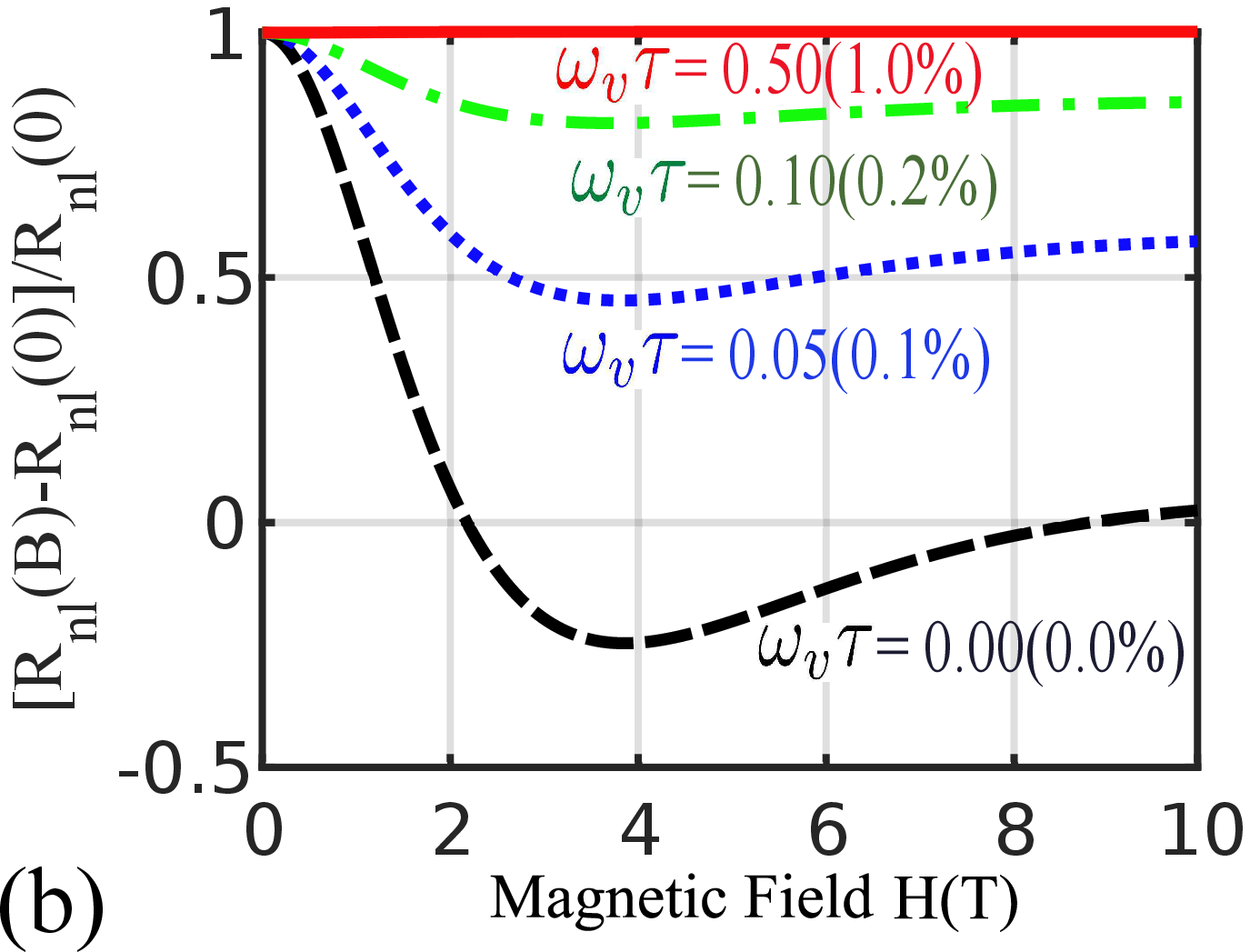}
\end{center}
\caption{(Color online) Nonlocal resistance $R_{nl}(H)$, in the
unit of $R_{nl}$(0), are plotted against magnetic field $H$ for different chemical potential (a) 
$\mu _{F}=0.15$ eV and (b) $\mu _{F}=-0.10$ eV.
Drude conductivity $\sigma _{D}$, Drude relaxation
time $\tau$ and  the 
scattering rate of spin, $\omega _{s}\tau$, can be obtained from the parameters of a microscopic scattering model\cite
{chunli2016graphene} : impurity density $n_{\mathrm{imp}}=5.0\times 10^{10}cm^{-2}$,
scalar potential $\mathcal{V}_{D}=50$ meV, SOC potential~ \cite{balakrishnan2014giant}  $\mathcal{V}_{S}=5$
meV, defect size $R=20$ nm, and associated momentum cutoff $k_c=2/R$. 
On the other hand, fairly modest strain can sustain
a large valley Hall effect~\cite
{zhang2016valley}, $\omega _{v}\tau (\gtrsim 1)$. $\omega _{v}\tau =1$ can be induced by applying along the
$y$ direction an average (uniaxial) strain of $2\%$.
Parameters: $\ell _{s}=0.53\protect\mu $m, $\ell _{v%
}=0.53\protect\mu $m, $w=0.50$$\protect\mu $m, $x=2.00\protect\mu$m and $y=0.25%
\protect\mu$m. 
}
\label{SMFIG2}
\end{figure}

In the presence of an in-plane magnetic field, the system response consists of eight types of currents. Recall that the magnetic field acts only as a Zeeman term that induces precession, and does not introduce a Lorentz force (i.e. $\vec{F}^B_{\vec{k}} = 0$ in Eq.~\eqref{QBE0}, as mentioned above). Accounting (phenomenologically) for spin relaxation,  the constitutive and continuity equations in the presence of the magnetic field read:
\begin{equation} \label{CE111111}
J_{i}^{\mu }=-(D_{ij})^{\mu}_{\nu }\partial _{j}N^{\nu }+(\sigma _{ij})^{\mu}_{\nu
}E_{j}^{\nu }, 
\end{equation}
\begin{equation} \label{3.399}
\partial_i J^{\mu}_i=-\frac{\delta ^{\mu}_{\nu}}{\tau^\nu}N^{\nu}+\omega^{\mu}_{\nu}
N^{\nu},
\end{equation}
where
\begin{equation}
D_{0}=
\begin{bmatrix}
D^0_{0} & 0 \\ 
0 & D^0_{0}  
\end{bmatrix},
D^0_{0}=\mathcal{D}_{r}%
\begin{bmatrix}
1 & \eta & 0 & 0 \\ 
\eta   & 1 & 0 & 0 \\ 
0 & 0 & 1 & \eta \\ 
0 & 0 & \eta & 1%
\end{bmatrix},
\end{equation}
\begin{equation}
D_{H}=
\begin{bmatrix}
0 & D^0_{H} \\ 
D^0_{H} & 0  
\end{bmatrix},
D^0_{H}=\mathcal{D}_{r}%
\begin{bmatrix}
\theta _{v} & \theta _{s} & 0 & 0 \\ 
\theta _{s} & \theta _{v} & 0 & 0 \\ 
0 & 0 & \theta _{v} & \theta _{s} \\ 
0 & 0 & \theta _{s} & \theta _{v}%
\end{bmatrix}.
\end{equation}
Substituting continuity equations \eqref{3.399} into the divergence of constitutive equations \eqref{CE111111}, the diffusion equations away from the boundaries take again a form similar to Eq.~\eqref{sdq},
\begin{equation} \label{2.5}
\partial_i ^{2}N^{\mu}-\mathcal{M}^{\mu}_{\nu }N^{\nu }=0.  
\end{equation}
However, this time the diffusion matrix is $4\times 4$ in order to accommodate the additional response modes introduced by the precession term:
\begin{equation}
\mathcal{M}\simeq%
\begin{bmatrix}
\ell _{v}^{-2} & -\eta \ell _{s}^{-2} & +\eta
\ell _{L}^{-2} & 0 \\ 
-\eta \ell _{v}^{-2} & \ell _{s}^{-2} & -\ell _{%
L}^{-2} & 0 \\ 
0 & +\ell _{L}^{-2} & \ell _{s}^{-2} & -\eta\ell _{%
v}^{-2} \\ 
0 & -\eta \ell _{L}^{-2} & -\eta \ell _{s}^{-2}
& \ell _{v}^{-2}%
\end{bmatrix}.
\end{equation}
The eigenvalues of the above diffusion matrix are
\begin{equation}
E_{\pm }^{\eta }=\frac{\ell _{v}^{-2}+\ell _{s}^{-2}\mp
i\ell _{L}^{-2}}{2}+\frac{\eta }{2}\sqrt{\Delta _{\pm }}.
\end{equation}
with 
\begin{equation}
\Delta _{\pm }=\left( \ell _{v}^{-2}-\ell _{s}^{-2}\pm
i\ell _{L}^{-2}\right) ^{2}+4\eta _{zz}^{2}\ell _{v%
}^{-2}\left( \ell _{s}^{-2}\mp i\ell _{L}^{-2}\right) .
\end{equation}
Hence, following the same procedure to find the solution as in the case with $H = 0$, we arrive at the following result for the NLR:
\begin{equation}
\frac{R_{nl}(x,H)}{R_{\mathrm{xx}}}=\frac{1}{\pi }\int_{-\infty
}^{\infty }dk\frac{e^{ikx}}{k}\frac{\tanh \left( k w/2\right) }{1+\sum_{b
}\Theta _{b }^{2}\mathcal{F}_{b}\left( k\right) },
\end{equation}
where we sum over four transverse eigenmodes $\{v,s,xo,yz\}$ in the denominator of the above integral. The above equation is the basis of the analysis about the suppression of the Hanle effect described in the main text.

\subsection{Carrier concentration dependence}

Fig.~\ref{SMFIG2} shows the NLR, $R_{nl}(x,H)$ normalized to its value at zero in-plane magnetic field, $R_{nl}(x,H=0)$ versus $H$, for different chemical potentials [(a) $\mu _{F}=0.15$ eV and (b) $\mu _{F}=-0.10$ eV]. As noticed in the main text, by setting $\omega _{v}\tau =0$, the result of Abanin \emph{el al.} \cite{abanin2009nonlocal} is recovered. In this  case, the diffusion lengths of the (spin) eigenmodes, $\ell_{s\pm }=(\ell _{s}^{-2}\pm i\ell _{%
L}^{-2})^{-1/2}$ become complex (with imaginary part $\ell _{L%
}^{-2}\propto H$), which leads to the development of an oscillatory component in the NLR (Hanle effect). Upon increasing the amount of nonuniform
strain, we find that the oscillating  part of the NLR is suppressed and even disappears for strains of the order of $\sim 1\%$. This shows that our result concerning the suppression of the Hanle effect for nonuniform strain of the order of a few percents maximum is robust against the change of the carrier density and sign.

\bibliography{reference}

\end{document}